\documentclass[12pt,a4paper]{article}

\usepackage{a4}          

\usepackage{amsmath}
\usepackage{amssymb}

\usepackage{hyperref}              
\hypersetup{pdftitle={Global F-theory GUTs}, 
    pdfauthor={Ralph Blumenhagen, Thomas W. Grimm, Benjamin Jurke and Timo Weigand}, 
    pdfsubject={String theory model building},
    bookmarksopen, bookmarksnumbered, bookmarksopenlevel=2}
    
\newcommand{\beq}{\begin{equation}}  \newcommand{\eeq}{\end{equation}}
\newcommand{\bal}{\begin{aligned}}   \newcommand{\eal}{\end{aligned}}
\newcommand{\bea}{\begin{eqnarray}}  \newcommand{\eea}{\end{eqnarray}}

\def\ov{\overline}

\newcommand{\ccP}{\mathcal{P}}

\newcommand{\bbC}{\mathbb{C}}
\newcommand{\bbP}{\mathbb{P}}

\newcommand{\GUT}{\mathrm{GUT}}
\newcommand{\surjto}{\to\!\!\!\!\!\to}

\newcommand{\tablelabel}[1]{\label{table:#1}}
\newcommand{\tableref}[1]{Table~\ref{table:#1}}

\newcommand{\seclabel}[1]{\label{sec:#1}}
\newcommand{\secref}[1]{section~\ref{sec:#1}}
\newcommand{\Secref}[1]{Section~\ref{sec:#1}}

\newcommand{\applabel}[1]{\label{app:#1}}
\newcommand{\appref}[1]{appendix~\ref{app:#1}}

\begin{document}

\baselineskip=14pt
\parskip 5pt plus 1pt

\vspace*{-1.5cm}
\begin{flushright}    
  {\small
  MPP-2009-148\\
  SLAC-PUB-13751
  }
\end{flushright}

\vspace{2cm}
\begin{center}        
  {\LARGE
  Global F-theory GUTs
  }
\end{center}

\vspace{0.75cm}
\begin{center}        
  Ralph Blumenhagen$^{1}$, Thomas W.~Grimm$^{2},$ \\[0.1cm]  
  Benjamin Jurke$^{1}$ and Timo Weigand$^{3}$
\end{center}

\vspace{0.15cm}
\begin{center}        
  \emph{$^{1}$ Max-Planck-Institut f\"ur Physik, F\"ohringer Ring 6, \\ 
               80805 M\"unchen, Germany } 
  \\[0.15cm] 
  \emph{$^{2}$ Bethe Center for Theoretical Physics and \\ 
               Physikalisches Institut der Universit\"at Bonn, Nussallee 12, \\ 
               53115 Bonn, Germany } 
  \\[0.15cm]
  \emph{$^{3}$ SLAC National Accelerator Laboratory, Stanford University, \\
               2575 Sand Hill Road, Menlo Park, CA 94025, USA }
\end{center}

\vspace{2cm}


\begin{abstract}
We construct global 
F-theory GUT models on del Pezzo surfaces in compact Calabi-Yau fourfolds realized as complete intersections of two hypersurface constraints. The intersections of the GUT brane and the flavour branes as well as the gauge flux are described by the spectral cover construction.
We consider a split $S[U(4)\times U(1)_X]$ spectral cover, which allows for the phenomenologically relevant Yukawa couplings and GUT breaking to the MSSM via hypercharge flux while preventing dimension-4 proton decay.
General expressions for the massless spectrum, consistency conditions and a new method for the computation of curvature-induced tadpoles are presented. We also provide a geometric toolkit for further model searches in the framework of toric geometry. Finally, an explicit global model with three chiral generations and all required Yukawa couplings is defined on a Calabi-Yau fourfold which is fibered over the del Pezzo transition of the Fano threefold $\bbP^4[4]$.
\end{abstract}

\clearpage


\tableofcontents

\newpage
\section{Introduction}

The success of the recently proposed $SU(5)$ grand unified models of particle physics in F-theory \cite{Donagi:2008ca, Beasley:2008dc, Beasley:2008kw, Donagi:2008kj} is largely rooted in the way how F-theory combines the main phenomenological achievements of model building in two different corners of the string landscape: the heterotic string and Type II strings with intersecting D-branes~\cite{Lust:2004ks, Blumenhagen:2005mu, Blumenhagen:2006ci, Nilles:2008gq}.

Just as for the $E_8 \times E_8$ heterotic string, the F-theory gauge dynamics is governed by the group theory of $E_8$. This is because F-theory provides a geometrization of Type IIB string compactifications with varying string coupling, and hence incorporates strong coupling enhancements to exceptional gauge groups. It has been known since the early days of the heterotic string that the gauge group and matter content of  GUT models, say the $SU(5)$ scenario, have a natural embedding into $E_8$. That the same is true even for the structure of Yukawa couplings is considered a major advantage of F-theory/heterotic GUT models over their weak-coupling Type II counterparts, which only realize the ${ \bf 10 \, 10 \, 5_H}$ Yukawa couplings non-perturbatively \cite{Blumenhagen:2007zk}.

Just as in perturbative Type II vacua with D-branes, the F-theory gauge dynamics is localized on subspaces of the internal manifold which encode the location of space-time filling seven-branes. This localization is substantially different from heterotic compactifications where the gauge dynamics arises from the closed string modes propagating on the entire internal space. Thus Type II/F-theory models in principle allow for  a local approach to string phenomenology which focuses on aspects of particle physics sensitive primarily to the geometry around the branes. All physics communicated through gravity interactions is decoupled at this stage of the analysis.

The very  existence of a local limit does not imply, however, that global aspects have become at any means irrelevant. For example, it is crucial to consider local models which admit a viable UV completion. This will pose strong conditions on the geometry in which the GUT seven-branes are embedded. In addition to these conditions on the geometry, one also needs to ensure that the scales in the problem are actually at the desired values to maintain consistency of a local approach. Since in string theory the scales are fixed by the vacuum expectation values of moduli fields, these questions must eventually be addressed in a full compact F-theory model. Moduli stabilization has been studied intensively in Type II/F-theory models \cite{Douglas:2006es, Denef:2008wq}, and it would be desirable to combine these scenarios with explicit realizations of realistic GUT models.

Concrete local $SU(5)$ GUT models have been studied by considering seven-branes on del Pezzo surfaces which are shrinkable inside the ambient space \cite{Donagi:2008ca, Beasley:2008dc, Beasley:2008kw, Donagi:2008kj}. Demanding the four-dimensional theory to be supersymmetric leads one to consider of F-theory on an elliptically fibered Calabi-Yau fourfold $Y$. The elliptic fiber corresponds to the varying complex dilaton-axion of Type IIB compactified on the base of this fourfold. The seven-branes are of complex codimension one in the three-dimensional base of $Y$ with the GUT brane wrapped on the del Pezzo surface. Matter arises from the intersection of the GUT brane with another matter seven-brane along complex curves. The mutual intersection points of two matter curves on the GUT brane in turn realize the Yukawa couplings between the charged fields. Phenomenological studies which can be performed just based on this ultra-local structure include refs.~\cite{Font:2008id, Heckman:2008qa, Blumenhagen:2008aw, Heckman:2009bi, Bouchard:2009bu, Heckman:2009de, Randall:2009dw, Tatar:2009jk, Conlon:2009qa, Font:2009gq}.

In order to address the even harder task to find global GUT models in F-theory on compact fourfolds, it is instructive to first approach this problem in the language of Type IIB orientifold compactifications \cite{Blumenhagen:2008zz}. In \cite{Blumenhagen:2008zz} the $SU(5)$ GUT brane is located on a pair of non-generic intersecting del Pezzo surfaces that arise via transitions from the famous quintic Calabi-Yau hypersurface in $\bbP^4$. Analyzing D7-brane fluxes and explicitly implementing the breaking of the GUT group to the Standard Model with a massless $U(1)_Y$, as proposed in \cite{Beasley:2008dc, Donagi:2008kj}, it was possible to precisely obtain the Standard Model spectrum. This also lead to the correct implementation of the intricate quantization conditions on the gauge flux \cite{Blumenhagen:2008zz}. In D-brane models, the $SU(5)$ on the GUT branes actually descends from a $U(5) = SU(5) \times U(1)_a$, and the diagonal $U(1)_a$ flux plays a crucial role in canceling the Freed-Witten anomalies. Also, avoiding exotic matter states in a way consistent with integer induced D3-charge of the flux requires a specific twist of the $U(1)_Y$ bundle with the $U(1)_a$ bundle. Finally, it is only with the help of this $U(1)_a$ that genuine gauge coupling unification without the introduction of a new threshold can be achieved \cite{Blumenhagen:2008aw}.

As we show in this work, genuine F-theory models can implement the successes of the global orientifold GUTs of \cite{Blumenhagen:2008zz} and also allow for the missing top Yukawa couplings. As an intermediate step to indicate a promising class of fourfold geometries, the lifting of the geometries of \cite{Blumenhagen:2008zz} to complete-intersecting Calabi-Yau fourfolds was considered in \cite{Blumenhagen:2009up} (see also \cite{Collinucci:2008zs, Collinucci:2009uh}). We will show that F-theory GUTs can in fact be realized on explicit Calabi-Yau fourfold geometries which are much simpler than the lifts \cite{Blumenhagen:2009up}, but share some crucial similarities in their local structure and construction.

In addition to the construction of compact Calabi-Yau fourfold geometries also the F-theory fluxes have to be incorporated to explicitly build viable GUT models. Recently, considerable progress was made towards a better understanding of the allowed gauge fluxes in F-theory. In \cite{Hayashi:2009ge, Donagi:2009ra} it was realized that a natural description of certain geometric aspects and the gauge flux is in terms of the spectral cover construction, which first came to stage in string theory in the context of heterotic compactifications \cite{Friedman:1997yq, donagi-1997-1}. Remarkably, these methods do not only apply to F-theory models which admit a heterotic dual, but appear to be valid for a much broader class of examples. In fact, this reasoning has been applied in refs.~\cite{Marsano:2009ym, Marsano:2009gv} to make progress in the construction of global models.

The aim of the present paper is to construct globally consistent F-theory GUT models. The Calabi-Yau fourfolds we will consider are elliptic fibrations over a special class of base threefolds $B$. We first pick a hypersurface base that admits an elliptic fibration which does not yield non-Abelian gauge groups in four dimensions, and then perform a transition to generate a del Pezzo surface $S$ in a new base $B$. The existence of the $\bf 10$ matter curve forces us to blow up a singular curve to a \textit{non-generic} del Pezzo surface. The Calabi-Yau fourfold $Y$ over the new $B$ is represented as the complete intersection of two hypersurface constraints in a six-dimensional ambient space. One of these constraints encodes the geometry of the base $B$ with the del Pezzo divisor $S$, while the second constraint contains the details of the elliptic fibration. By degenerating the elliptic fiber over $S$ an $SU(5)$ gauge group can be generated on the GUT seven-brane. These steps can be made explicit using methods of toric geometry, as we show for a del Pezzo $7$ transition of the quatric hypersurface in~$\bbP^4$. 

A key feature of F-theory is that the non-Abelian enhancement along the GUT brane renders the Calabi-Yau fourfold singular. These singularities have to be resolved when analyzing the geometry of the GUT model fourfold $Y$. In particular, this will alter the Euler characteristic $\chi(Y)$ which measures the curvature induced 3-brane charge. For an $SU(5)$ GUT with further gauge enhancement over curves and points, $\chi(Y)$ exemplifies a considerable departure from the naive value for the fourfold with no non-Abelian enhancements. Remarkably, this departure can be computed not only by direct geometric methods, such as explicit toric blow-ups or the method of  \cite{Andreas:1999ng, Andreas:2009uf}, but also by employing the auxiliary spectral cover construction. 

The spectral cover construction also allows us to encode the gauge flux bundles. As one of our findings we stress the importance of bundles with structure group $S[U(N) \times U(1)]$ in the context of $SU(5)$ GUT models. First, as in the Type IIB context, the consistent breaking of $SU(5)$ via $U(1)_Y$ flux requires a twisting of the hyperflux bundles with another bundle with non-vanishing first Chern class. This means that in the spectral cover approach the gauge flux on the matter branes is actually not described by an $SU(5)$ bundle, but by a $U(5)$ bundle related to the hyperflux bundle in a specific manner. This construction was in fact worked out in the context of the heterotic string in \cite{Blumenhagen:2005ga, Blumenhagen:2006ux, Weigand:2006yj} to incorporate the GUT breaking in exactly the manner that has come to fame in F-theory models since the appearance of \cite{Beasley:2008dc, Donagi:2008kj}. 

But the relevance of the $U(N)$ bundles is not restricted to GUT symmetry breaking. In \cite{Tatar:2009jk} it was found that absence of unacceptable dimension four proton decay operators
requires that the naive $SU(5)_{\perp}$ spectral cover describing the geometry transverse to the GUT brane has to factorize.  The resulting massive $U(1)$ symmetry then gives the otherwise missing  selection rules for the couplings (see also \cite{Tatar:2006dc}). This construction was further worked out in \cite{Marsano:2009gv}. Their analysis singles out  as the preferred such splitting the factorization of the spectral cover into a rank four and a rank one cover.  We identify the associated gauge bundles  as of type $S[U(4) \times U(1)_X]$.  This class of embeddings had likewise come to considerable use in heterotic GUT model building in \cite{Blumenhagen:2005ga, Blumenhagen:2006ux, Blumenhagen:2006wj, Weigand:2006yj}. The bundles are subject to subtle quantization conditions. From a practical perspective this construction bypasses the no-go theorem of \cite{Donagi:2009ra} for the realization of chiral three generations with so-called universal gauge flux in models on generic del Pezzo surfaces. On the other hand it is now necessary to obey the D-term constraint  associated with the massive $U(1)_{X}$ symmetry.

These considerations eventually enable us to present an explicit $SU(5)$ GUT model with three chiral generations of ${\bf 10}$ and ${\bf \ov 5_m }$ on our compact Calabi-Yau fourfold.  The model indeed realizes the ${\bf 10 \, 10 \, 5_H}$ and ${\bf 10 \, \ov 5_m \, \ov 5_H}$ Yukawas, while the dimension 4 proton decay operators are absent by virtue of the split spectral cover.  Even though a detailed study of the phenomenological properties of this example is beyond our scope we note that the model does contain candidates for right-handed neutrinos and comment
on the issue of gauge coupling unification in such F-theory models. The D-term supersymmetry conditions can be met inside the K\"ahler cone, but the 3-brane tadpole of the flux overshoots the curvature part.

\subsubsection*{Guide through the paper}
The paper is organized as follows. In \secref{CompactGeometries} we introduce a general class of complete-intersecting Calabi-Yau fourfolds which are candidates to support simple GUT models. We outline in \secref{delPezzo} how the base of an elliptically fibered fourfold is obtained by a del Pezzo transition. The complete-intersecting Calabi-Yau fourfold and its elliptic fibration structure are discussed in \secref{complete_intersect_gen}. In \secref{GeomTad} we then investigate the geometric tadpoles and present a simple expression for the Euler characteristic $\chi(Y)$ for singular fourfolds. The details of this computation are delegated to an appendix.

In \secref{gen_SU(5)GUT} we introduce the details of F-theory compactifications leading to an $SU(5)$ GUT model. First, in \secref{SU(5)features}, we recall how the degenerations of the elliptic fibration yield the required matter representations and Yukawa couplings. We also argue that the presence of a $\bf{10}$ representation restricts the GUT brane del Pezzo surface to be non-generic and shrinkable only to a singular curve. In \secref{spectral_cover}, we introduce the spectral covers, and describe the geometry of the $SU(5)$ GUT geometry within the auxiliary manifold associated with the spectral construction. The gauge fluxes are introduced in \secref{gauge_fluxSU(5)}, where we also present the concrete expressions for the chiral index of the respective matter curves. The GUT breaking by hypercharge flux and the flux induced D3-tadpole are discussed in \secref{hyper}.

\Secref{sec_split} is devoted to a discussion of the $S[U(4)\times U(1)_X]$ construction on which our explicit model will be based. After reviewing in \secref{split_covers} the factorization of the spectral cover, \secref{sec_Bundle} describes in detail the $S[U(4)\times U(1)_X]$  embedding governing the gauge bundle. Special emphasis is put on the quantization of the gauge flux and the computation of the chiral matter content. \Secref{split_GUTbreaking} explains the GUT symmetry breaking via $U(1)_Y$ hyperflux as a specific twisting of the gauge bundles and analyzes the resulting 3-brane tadpole. The supersymmetry conditions are the subject of \secref{SUSY_sec}.

Finally in \secref{sec_geometry} we present a concrete Calabi-Yau fourfold example $Y$ on which we realize a three generation GUT model. This will allow us to include a brief introduction to the relevant toric tools to explicitly construct the complete-intersecting fourfold $Y$. In \secref{FanodP7trans} we first study the del Pezzo transition of a simple Fano threefold, the quartic in $\bbP^4$. Using toric geometry we construct a threefold base $B$ by blowing up a singular $\bbP^1$ into a del Pezzo $7$ surface. In \secref{complete_inters_ex} we then realize the elliptic fibration over $B$ as a complete-intersecting Calabi-Yau fourfold in a six-dimensional toric ambient space. The elliptic fibration can be degenerated to $SU(5)$ over the del Pezzo surface. Again applying toric techniques, we are able to resolve this singular space and explicitly compute the induced geometric 3-brane tadpole given by $\chi(Y)$ in \secref{toric_resolution_ex}. Finally, in \secref{sec_example}, we construct an $S[U(4)\times U(1)_X]$ split spectral cover for the GUT and matter branes which encodes gauge flux for a three generation GUT model. We also comment on phenomenological aspects such as gauge coupling unification and the potential reconciliation of neutrino physics with the appearance of a split spectral cover, before presenting our conclusions in \secref{sec_concl}.

\section{Compact geometries for GUT model building}\seclabel{CompactGeometries}

The physics of F-theory compactifications is encoded in two pieces of data: the geometry of an in general singular elliptically-fibered Calabi-Yau fourfold $Y$ as well as the gauge flux on the seven-branes, described in terms of the background value of the four-form flux $G$. In this section we introduce the compact Calabi-Yau fourfolds on which we will build explicit GUT models in the remainder of this work. This will allow us to comment on the main technology to construct and study geometries which permit $SU(5)$ GUT models.

\subsection{Calabi-Yau fourfold bases and del Pezzo transitions}\seclabel{delPezzo}
F-theory on an elliptically fibered Calabi-Yau fourfold $Y$ with base $B$ is equivalent to Type IIB string theory on $B$ with a dilaton-axion $\tau=C_0 + \mathrm{i} e^{-\phi}$ varying over this base. At each point in $B$ the complex number $\tau$ can be identified with the complex structure modulus of the elliptic fiber over this point. For $Y$ to be a Calabi-Yau fourfold this fiber has to degenerate over divisors $D_i$ in $B$. These degeneration loci encode the location of space-time filling seven-branes of Type IIB compactified on $B$ \cite{Denef:2008wq}. 

Only a very special class of Calabi-Yau fourfolds admit global elliptic fibrations which do not render the Calabi-Yau manifold singular. As a sufficient criterion to ensure the existence of such a smooth elliptic Calabi-Yau manifold as fibration over $B$ one can require $-K_B$ to be very ample, which roughly speaking means that $B$ admits an embedding into projective space \cite{Hartshorne:1977AlgGeom}.\footnote{The base also has to be free of irregularities, i.e.~$h^i(B;{\cal O}_B)=0$ for all $i>0$.} A prominent class of such base spaces $B$ can be found among the smooth Fano threefolds which have been classified by Iskovskih and Mori-Mukai \cite{Itskovskih1, Itskovskih2, MoriMukai1, MoriMukai2}. There are only a finite number of smooth Fano threefolds and their geometry has been studied intensively in the mathematical literature. To find a list of the subclass of very ample Fano threefolds we refer the reader to ref.~\cite{Grassi:1997mr}. Crucial for us will be two facts about these base geometries:

\begin{itemize}
  \item[1.] Since the elliptically fibered fourfolds over $B$ are generically non-singular there will be \textit{no} non-Abelian gauge groups in the four-dimensional effective theory. The Euler characteristic of such fourfolds $Y$ is simply given by \cite{Sethi:1996es, Klemm:1996ts}
    \beq \label{smooth_chi}
      \chi(Y) = 360 \int_B c_1^3(B)+ 12 \int_B c_1(B)\, c_2(B) ,
    \eeq
    where $c_i(B)$ are the Chern classes of $B$. 
  \item[2.] Some of these very ample Fano threefolds can be represented as simple hypersurfaces in a weighted projective or toric ambient space. Our key example will be the quartic in $\bbP^4$ which is given by a polynomial of degree four in $\bbP^4$ and will be denoted by $\bbP^4[4]$. Other simple examples are $\bbP^4[2]$ and $\bbP^4[3]$. In general, each such hypersurface is given by a polynomial constraint
    \beq \label{Fano_base}
      P_{\rm base}(y_i) = 0 ,
    \eeq
    where $y_i$ are the coordinates of the projective or toric ambient space and are related by a number of scaling relations.
\end{itemize}

We are aiming to build GUT models on seven-branes wrapped on a special class of divisors, namely del Pezzo surfaces, in the base $B$. In fact, the del Pezzo surfaces are by definition precisely the two-dimensional Fano manifolds. In working with Fano threefolds it is thus instructive to compare this with the more familiar situation in two dimensions. The $10$ del Pezzo surfaces are $\mathbb{P}^1 \times \mathbb{P}^1$ and $dP_i$, which is $\mathbb{P}^2$ with $n=0,\ldots,8$ points blown up to $\mathbb{P}^1$'s. Note that the del Pezzo surfaces $\mathbb{P}^1 \times \mathbb{P}^1$ and $dP_n,\, n=1,2,3$ are two-dimensional toric varieties, while the remaining del Pezzo surfaces can be represented as hypersurfaces or complete intersections in a higher-dimensional toric ambient space. In particular, three of these del Pezzo surfaces can be represented as hypersurfaces in weighted projective space as
\beq \label{dP678_hyper}
  dP_6 = \mathbb{P}_{1111}[3] , \qquad dP_7 = \mathbb{P}_{1112}[4] , \qquad dP_8 = \mathbb{P}_{1123}[6] ,
\eeq
where the subscripts denote the weight of the $4$ projective coordinates and we have also indicated the degree of the hypersurface in the square brackets. These are the analogues of the Fano threefold hypersurfaces discussed in the second point above. The rich geometry of the del Pezzos \eqref{dP678_hyper} allows for flexible model building.
Our strategy is to generate the del Pezzo surfaces in a base $B$ by performing del Pezzo transitions starting with a Fano threefold. In order to generate the hypersurface del Pezzos \eqref{dP678_hyper}, the most direct approach is to start with a hypersurface Fano threefold defined by \eqref{Fano_base}.

To discuss the del Pezzo transitions we first note that this analysis is similar to del Pezzo transitions in Calabi-Yau threefolds as discussed, for example, in refs.~\cite{Grimm:2008ed, Blumenhagen:2008zz}. In a first step one has to render the manifold $B$ singular and generate a del Pezzo singularity. This implies that one has to fix a number of complex structure moduli, which appear as the coefficients in \eqref{Fano_base}. As a simple example one can generate a $dP_6$ singularity in the Fano threefold $\bbP^4[4]$ by tuning the defining polynomial to take the form 
\beq
  P_{\rm base}(y_i) = y_5 f_{3}(y_1,y_2,y_3,y_4)+ f_{4}(y_1,y_2,y_3,y_4) , 
\eeq 
where $f_n$ are polynomials of degree $n$ which only depend on the four coordinates $(y_1,y_2,y_3,y_4)$. The polynomial $f_3$ precisely parametrizes the del Pezzo $6$ singularity at the point $(y_i) = (0,0,0,0,1)$. One can then blow up the singularity by pasting in a $dP_6$ surface $S$ of finite size. After the del Pezzo transition the new base $B$ will be parametrized by one more coordinate $w$ as
\beq
  P_{\rm base}(y_i,w) = y_5\, f_{3}(y_1,y_2,y_3,y_4)+ w\, f_{4}(y_1,y_2,y_3,y_4) ,
\eeq 
and thus admit an additional exceptional $dP_6$ divisor $S$ given by $w=0$. This provides a very simple example in which a point-like singularity is blown up into a generic del Pezzo surface. This generic del Pezzo surface is given by a completely generic cubic equation in~$\bbP^3$. However, as we will discuss in \secref{SU(5)features} these generic del Pezzos are \textit{not} good candidates for GUT model building. One thus has to employ a more involved geometric transition, which we want to describe next.

Instead of blowing up points to del Pezzo surfaces these can also be generated from more complicated singular loci such as complex curves. In these cases the geometric transitions will generate non-generic del Pezzo surfaces. In the example which we will study intensively in this work we will generate a non-generic $dP_7$ surface in the Fano threefold $\bbP^4[4]$ by blowing up a singular $\bbP^1$. In principle the strategy is similar to the one in the generic $dP_6$ case just discussed. One first tunes the coefficients of the quatric in $\bbP^4$ such that one obtains a singularity along a $\bbP^1$ parametrized by $(0,0,0,y_4,y_5) \sim \lambda (0,0,0,y_4,y_5)$ for $\lambda \in \bbC^\times$ which can be blown up into a finite size $dP_7$. The blowup is again obtained by including a new coordinate $w$ and an additional scaling relation. The transitioned base manifold $B$ now has one additional divisor $S$ which again is given by $w=0$. This $S=dP_7$ is now non-generic, and we will see in \secref{sec_example} that it takes the form 
\beq
  P_{dP_7} = p^2 + f_2(y_1,y_2,y_3) g_2(y_1,y_2,y_3) ,
\eeq
where $f_2$ and $g_2$ are generic polynomials of degree $2$ in $(y_1,y_2,y_3)$, and the coordinates $(p,y_1,y_2,y_3)$ have scaling weight $(1,1,1,2)$ as required in \eqref{dP678_hyper}. We will discuss this example in detail in \secref{sec_example} and show that it is a simple candidate to host an $SU(5)$ GUT model. Let us stress, that typically for all Fano base manifolds there are numerous del Pezzo transitions of this type. They are best studied using toric geometry which also allows us to compute the relevant topological data such as the intersection numbers and Chern classes of~$B$. 

Let us summarize our strategy. We have started with an elliptic fibration over a Fano hypersurface which left us with no non-Abelian gauge group in the four-dimensional effective theory. Then we performed a del Pezzo transition to a new base $B$. On this del Pezzo surface we want to place our GUT seven-brane, which amounts to degenerate the elliptically fibration of the Calabi-Yau fourfold over this del Pezzo to an $SU(5)$ singularity.

\subsection{Complete-intersecting Calabi-Yau fourfolds}\seclabel{complete_intersect_gen}
So far we have mainly discussed the base $B$ of the Calabi-Yau fourfold $Y$. To explicitly realize $Y$ itself one needs to also describe the structure of the elliptic fibration. This is of particular importance since the degeneration of the elliptic fiber encodes the physics on the world-volume of the seven-branes. The Calabi-Yau fourfolds we will consider are thus given by \textit{two} hypersurface constraints 
\beq \label{two_constr}
   P_{\rm base}(y_i,w) = 0 ,\qquad P_{\rm W}(x,y,z,y_i,w)=0 ,
\eeq
in a six-dimensional projective or toric ambient space. Here $P_{\rm W}$ is the Weierstrass form \eqref{Tate1} encoding the structure of the elliptic fibration. Most generically the constraint $P_{\rm W}=0$ can be brought into the Tate form 
\beq \label{Tate1}
    P_{\rm W} = x^3 - y^2 + x\, y\,  z\, a_1 + x^2\, z^2\, a_2 + y\, z^3\,a_3   + x\, z^4\, a_4 + \, z^6\, a_6\ = 0,
\eeq
where $(x,y,z)$ are coordinates of the torus fiber. In the sequel we will only be working with the inhomogeneous Tate form by setting $z=1$. The $a_n(y_i,w)$ are sections of $K_B^{-n}$, with $K_B$ being the canonical bundle of the base $B$.  Thus the $a_n$ depend on the complex coordinates $y_i,w$ of the base $B$. Setting all $a_n=1$ one finds that \eqref{Tate1} reduces to the elliptic fiber $\bbP_{123}[6]$.

The $a_n$ encode the discriminant of the elliptic fibration. One first introduces the new sections 
\beq
  \beta_2 = a_1^2 + 4 a_2 ,\qquad 
  \beta_4 = a_1 a_3 + 2\, a_4 ,\qquad
  \beta_6 = a_3^2 + 4 a_6  ,
\eeq
where $\beta_i\in H^0(B;K_B^{-i})$. The discriminant can then be expressed as
\beq
  \Delta = -\tfrac14 \beta_2^2 (\beta_2 \beta_6 - \beta_4^2) - 8 \beta_4^3 - 27 \beta_6^2 + 9 \beta_2 \beta_4 \beta_6 ,
\eeq
which is a section of $K_B^{-12}$. In general, the discriminant $\Delta$ will factorize with each factor describing the location of a 7-brane on a divisor $D_i$ in $B$. Let us denote by $\delta_i$ the vanishing degree of the discriminant $\Delta$ over the divisor $D_i$. For higher degenerations this will also introduce non-trivial gauge-groups on the 7-branes. The precise group is encoded by the vanishing degree of the $a_i$ and $\Delta$ as classified in Table~2 of ref.~\cite{Bershadsky:1996nh}. For example, as we will explain in greater detail in \secref{SU(5)features}, for an $SU(5)$ gauge group along the divisor $w=0$ the sections $a_i$ have to take the form
\beq \label{TateSU5b1}
  a_1 = \mathfrak{b}_5 , \quad
  a_2 = \mathfrak{b}_4 w , \quad
  a_3 = \mathfrak{b}_3 w^2 , \quad
  a_4 = \mathfrak{b}_2 w^3 , \quad
  a_6 = \mathfrak{b}_0 w^5 ,
\eeq
where the sections $\mathfrak{b}_i$ generically depend on all coordinates $(y_i,w)$ of the base $B$ but do not contain an overall factor of $w$. It is important to stress that in case of such a higher degeneration, not only the elliptic fibration will be singular, but rather the Calabi-Yau fourfold itself. 

The compact Calabi-Yau fourfolds explicitly realized via \eqref{two_constr} as complete intersection in a projective or toric ambient space can be studied using toric geometry as we exemplify for our example in \secref{sec_example}. Note that in case $Y$ is still smooth, i.e.~in the absence of non-Abelian gauge groups as in \secref{delPezzo}, one can directly use toric geometry to compute the relevant topological data. However, the GUT models we will consider in \secref{gen_SU(5)GUT} have $SU(5)$ gauge groups on the blown up del Pezzo divisor $S$ in the base. The $a_i$ thus have to take the non-generic form \eqref{TateSU5b1} such that the Calabi-Yau fourfold $Y$ degenerates appropriately over $S$. In order to nevertheless determine the topological data of the $SU(5)$ GUT fourfold $Y$ one has to resolve these singularities. In case one has a concrete realization \eqref{two_constr} within a toric framework, one can do this explicitly as we show for our example in \secref{toric_resolution_ex}.

In our GUT models we will have only a non-Abelian gauge symmetry on the del Pezzo divisor $S$. It turns out that 
this allows us to use another set of powerful tools to analyze the geometry. Namely, we will apply the spectral cover constructions familiar from heterotic F-theory duality to construct our GUT models. It turns out that these yield very elegant and concise results which pass non-trivial tests when compared with the toric analysis. This includes a simple equation for the Euler characteristic of the singular Calabi-Yau fourfold $Y$ which we will summarize next, and derive in \appref{Tadpoles_sec}.

\subsection{Geometric tadpoles}\seclabel{GeomTad}
As in a perturbative Type IIB compactification with D-branes and fluxes, the cancellation of tadpoles is of crucial importance to define a global F-theory vacuum. In general, there will be induced tadpoles from 7-brane, 5-brane and 3-brane charges. Of these only the 7-brane and 3-brane tadpole consist in or, respectively, contain pieces induced entirely by the geometry of the fourfold compactification. It is these purely curvature dependent contributions that we will discuss in the following.

The seven-brane tadpole is automatically satisfied in F-theory provided the compact fourfold $Y$ is a Calabi-Yau manifold. In fact for an elliptic fibration the first Chern class of the base must be related to the discriminant locus as
\beq \label{discr_a}
  \sum_i \delta_i \, D_i = 12 \, c_1(B) ,
\eeq
where the $D_i$ are the divisors in the base $B$ over which the fiber degenerates with degree $\delta_i$ in the discriminant. The Tate classification of these vanishing degrees for different gauge groups can be found, for example, in Table~2 of ref.~\cite{Bershadsky:1996nh}. 

A non-trivial compact geometry of $Y$ also induces a non-vanishing contribution to the three-brane tadpole. The full cancellation condition is of the form~\cite{Sethi:1996es}
\beq
\label{D3a}
   \frac{\chi(Y)}{24} = N_{3} + \frac12 \int_Y G \wedge G . 
\eeq
Here $N_{3}$ is the charge of space-time filling 3-branes which are points in the fourfold $Y$, and $G$ is the four-form flux on $Y$ which we will discuss in more detail below. The presence of $G$ is crucial since it encodes the two-form gauge fluxes on the 7-branes which allow for the presence of chiral matter. The challenge is thus to find compactifications with a sufficiently large $\chi(Y)$ to allow for a non-negative $N_{3}$ and a non-vanishing $G$ supporting a GUT model. It is thus important to compute the Euler characteristic $\chi(Y)$ for a fourfold that gives rise to a non-Abelian gauge symmetry in four space-time dimensions.

Recall that in the presence of a non-Abelian gauge symmetry the fourfold $Y$ is a singular space. The Euler character $\chi(Y)$ therefore actually refers to a suitable resolution of this singular space. As we will see, this value is generically reduced drastically compared to the expression for a non-singular elliptic fibration. In this work we are using three complementary methods to compute $\chi(Y)$ for a given Weierstrass model.
\begin{itemize}
  \item[1.] In \secref{sec_geometry} we explicitly construct the resolution of the singular fourfold and compute its Euler characteristic with toric methods. This technique can be applied to a very large class of models where the singular fibers can be resolved by extending the toric ambient space.
  \item[2.] For general elliptic fibrations, a mechanism to compute the effect of the singularities was developed in \cite{Andreas:1999ng, Andreas:2009uf}. We have applied and slightly extended this method to check our results. However, since this analysis is rather involved we do not present the details here. Instead we have chosen to present yet another method which gives a closed expression for $\chi(Y)$ inspired by heterotic- F-theory duality.
  \item[3.] In case a heterotic dual exists for the F-theory compactification one can use the fact that the 3-brane charge has to match the number of heterotic M5-branes to derive a closed expression for $\chi(Y)$ as shown in \appref{Tadpoles_sec}. Remarkably, this expression also computes the correct Euler number for fourfolds which do not necessarily posses a heterotic dual, but admit only one gauge-group over the divisor $S$ as constructed in \secref{delPezzo}.  
\end{itemize}

\begin{table}[ht] 
  \renewcommand{\arraystretch}{1.3} 
  \centering
  \begin{tabular}{c|c|c} 
    $H = E_8/G$ & $G$ & $\chi_G$$_\big.$      \\ 
    \hline\hline 
    $E_{9-n}, \ n\le 5$ & $SU(n)$ &  $\int_S c_1^2(S) (n^3-n) +3n\, \eta \big(\eta-n c_1(S)\big)$$^\big.$ \\
    $SU(3)$    & $E_6$ & $72 \int_S  \bigl( \eta^2 - 7\eta c_1(S) + 13 c_1^2(S)\bigr)$ \\
    $SU(2)$  &  $E_7$ & $18 \int_S \bigl( 8\eta^2 - 64\eta c_1(S) + 133 c_1^2(S)\bigr)$ \\
    - & $E_8$ & $120 \int_S  \bigl( 3\eta^2 - 27\eta c_1(S) + 62 c_1^2(S)\bigr)$ \\
  \end{tabular} 
  \caption{\small Redefined Euler characteristic for $E_n$-type gauge groups. Here $\eta$ is given by $\eta =6c_1(S) + c_1(N_S)$.}
  \tablelabel{chigaugegroups}
\end{table} 

Let us now summarize the equation which captures $\chi(Y)$ for gauge group $H=E_8/G$ so that $G$ is the complement of $G$ in $E_8$. We will thus consider a Calabi-Yau fourfold with an elliptic fibration \eqref{Tate1} which degenerates to yield the four-dimensional gauge group $H$ from some 7-brane divisor $S$, but otherwise induces no further four-dimensional gauge factors. This is the case for the base spaces $B$ constructed from Fano threefolds as in \secref{delPezzo}. The Euler characteristic is then of the form  
\beq \label{chi-prop}
  \chi(Y) = \chi^*(Y) + \chi_{G} - \chi_{E_8} ,
\eeq
where $\chi^*(Y)$ is the Euler characteristic for a smooth fibration on $B$ given in \eqref{smooth_chi}. The Euler characteristic $\chi_G$ for the gauge groups $H=E_8/G$ and $\chi_{E_8}$ are listed in \tableref{chigaugegroups}. If the complement $G$ of $H$ in $E_8$ splits into the product of two groups the part $\chi_G$ is replaced by the corresponding sums.

Our conjecture is that the expression \eqref{chi-prop} holds for every Calabi-Yau fourfold $Y$ which is elliptically fibered over a base $B$ that also allows a generic Weierstrass model whose discriminant consists only of $\mathrm{I}_1$ or $\mathrm{II}$  components but contains no loci of further non-Abelian enhancements. We have checked this for a number of examples using the methods 1. and 2. listed above. The conjectural part of this assertion is non-trivial for the following reason: For a generic compact model the coefficients ${\mathfrak{b}}_i$ in \eqref{TateSU5b1} and their generalizations do depend also on the coordinate $w$ normal to $S$ at subleading order, and in general it is not possible to set these terms to zero without rendering the manifold unacceptably singular. By contrast, in our formula the sections $\eta$ appearing in \tableref{chigaugegroups} are entirely the pullback of sections on $S$. The consistency check gives us confidence, though, that this discrepancy is irrelevant for the computation of the Euler characteristic. It would be interesting to understand the deeper reason for this match.

\section{F-theory compactifications with \boldmath$SU(5)$ gauge symmetry}\seclabel{gen_SU(5)GUT}

In this section we focus concretely on the general construction of $SU(5)$ GUT models on del Pezzo surfaces. We first introduce the general degenerations of the Tate form \eqref{Tate1} and discuss the allowed matter representations and couplings. Then we introduce the spectral cover construction in \secref{spectral_cover} which will be of particular importance in \secref{gauge_fluxSU(5)} in the analysis of the seven-brane world-volume fluxes. Finally, in \secref{hyper} we discuss the breaking of the GUT group and compute the 3-brane tadpole induced by a suitably twisted hypercharge flux.

\subsection[$SU(5)$ GUT models]{\boldmath$SU(5)$ GUT models}\seclabel{SU(5)features}
In \secref{complete_intersect_gen} we introduced the general Tate form \eqref{Tate1} of an elliptically fibered Calabi-Yau fourfold. We recalled that the degeneration of the elliptic fibration over divisors in the base encodes the gauge group in the four-dimensional effective theory. Let us now specialize to the situation with an $SU(5)$ singularity along the GUT divisor $S \subset B$ given by the equation 
\beq
  S: \quad w =0  .
\eeq
Using Tate's algorithm this occurs if the complex structure moduli are tuned such that the coefficients of \eqref{Tate1} take the form 
\beq \label{TateSU5b}
  a_1 = \mathfrak{b}_5 , \quad
  a_2 = \mathfrak{b}_4 w , \quad
  a_3 = \mathfrak{b}_3 w^2 , \quad
  a_4 = \mathfrak{b}_2 w^3 , \quad
  a_6 = \mathfrak{b}_0 w^5 ,
\eeq
where the sections $\mathfrak{b}_i$ generically depend on all coordinates $(y_i,w)$ of the base $B$.

It is straightforward to evaluate the discriminant $\Delta$ of the elliptic fibration in terms of the new sections $\mathfrak{b}_i$ as
\beq \label{Delta_SU5}
  \Delta =  - w^5 \, \left(  \mathfrak{b}_5^4  P  + w  \mathfrak{b}_5^2 (  8  \mathfrak{b}_4 P +  \mathfrak{b}_5  R  ) + w^2 (16  \mathfrak{b}_3^2  \mathfrak{b}_4^2 +  \mathfrak{b}_5 Q) + {\cal O}(w^3) \right)
\eeq
with
\beq
  P = \mathfrak{b}_3^2 \mathfrak{b}_4 - \mathfrak{b}_2 \mathfrak{b}_3 \mathfrak{b}_5 + \mathfrak{b}_0 \mathfrak{b}_5^2 , \qquad
  R = 4 \mathfrak{b}_0 \mathfrak{b}_4 \mathfrak{b}_5- {\mathfrak{b}_3^3} - \mathfrak{b}_2^2 \mathfrak{b}_5 .
\eeq
Generically the expression in brackets in \eqref{Delta_SU5}, denoted as $D_1$, does not factorize further and thus constitutes the single-component locus of an $\mathrm{I}_1$ singularity. 
Cohomologically, one thus finds that the class $[\Delta]$ splits as $[\Delta] = 5 [S] + [D_1]$.

\begin{table}[t]
  \centering
    \begin{tabular}{lc|c|cc|ccccc|ll}
                 &   sing.                        & discr.         & \multicolumn{2}{|c}{gauge enh.}        & \multicolumn{5}{|c}{coeff.~vanish.~deg} & \multicolumn{2}{|c}{object}  \\
                 &   type                         & $\deg(\Delta)$ & type & group  & $a_1$ & $a_2$ & $a_3$ & $a_4$ & $a_6$${}_\big.$ & \multicolumn{2}{|c}{equation} \\ \hline\hline
      GUT:       & $\mathrm{I}_5^{\,\mathrm{s}}$  & 5 & $A_4$ & $SU(5)$    & 0 & 1 & 2 & 3 & 5 & $S:$ & $w=0$${}_\big.$${}^\big.$ \\
      matter:    & $\mathrm{I}_6^{\,\mathrm{s}}$  & 6 & $A_5$ & $SU(6)$    & 0 & 1 & 3 & 3 & 6 & $P_5:$ & $P=0$ \\
                 & $\mathrm{I}_1^{*\,\mathrm{s}}$ & 7 & $D_5$ & $SO(10)$   & 1 & 1 & 2 & 3 & 5 & $P_{10}:$ & $b_5=0$${}_\big.$ \\
      Yukawa:{\!\!\!\!\!} & $\mathrm{I}_2^{*\,\mathrm{s}}$ & 8 & $D_6$ & {\!\!\!}$SO(12)^*$ & 1 & 1 & 3   & 3 & 5 & \multicolumn{2}{|l}{$\mathfrak{b}_3=\mathfrak{b}_5=0$} \\
                 & $\mathrm{IV}^{*\,\mathrm{s}}$  & 8 & $E_6$ & $E_6$      & 1 & 2 & 2 & 3 & 5 & \multicolumn{2}{|l}{$\mathfrak{b}_4=\mathfrak{b}_5=0$}${}_\big.$ \\
      extra:     & $\mathrm{I}_7^{\,\mathrm{s}}$  & 7 & $A_6$ & $SU(7)$    & 0 & 1 & 3 & 4 & 7 & \multicolumn{2}{|l}{$P=R=0,$} \\
                 &                                &   &       &            &   &   &   &   &   & \multicolumn{2}{|l}{$(\mathfrak{b}_4,\mathfrak{b}_5)\not=(0,0)$}
    \end{tabular}
  \caption{\small Relevant gange enhancements in the considered $SU(5)$ GUT geometry.} 
  \tablelabel{tab:GaugeEnhanc}
\end{table}

The singularity structure enhances further along certain subloci on $S$. Moreover codimension-2 enhancements (as viewed on $B$) give rise to the so-called matter curves along which zero modes charged under the $SU(5)$ gauge group are localized. Matter in the ${\bf 10}$ representation is hosted by the curve of $SO(10)$ enhancement
\beq \label{curve10}
  P_{10}: \quad w=0 \quad \cap \quad \mathfrak{b}_5 = 0.
\eeq
along which the discriminant \eqref{Delta_SU5} scales like $w^7$, as required for a $D_5$ singularity. Matter in the ${\bf 5}$ lives on the  $SU(6)$ enhancement locus
\beq \label{curve5a}
  P_5: \quad w=0 \quad \cap \quad P = \mathfrak{b}_3^2 \mathfrak{b}_4 - \mathfrak{b}_2 \mathfrak{b}_3 \mathfrak{b}_5 + \mathfrak{b}_0 \mathfrak{b}_5^2 = 0,
\eeq
consistent with the scaling $\Delta \propto w^6$ in \eqref{Delta_SU5}. Note that for generic choice of sections $\mathfrak{b}_i$ the curve $P_5$ does not factorize so that all matter in the fundamental representation, both ${\bf \ov 5}_m$ and the Higgs ${\bf 5_H} + {\bf \ov 5_H}$ are localized on the same curve. As we will recall momentarily this is not desirable for phenomenological reasons.

The singularity enhances even further in codimension-3, i.e. at singular points, which  encode the existence of Yukawa couplings among the matter fields. There are three types of such points of higher enhancement in the present context:
\begin{itemize}
  \item ${\bf 10 \, 10 \, 5_H}$ Yukawa is localized at  point of $E_6$ enhancement $\mathfrak{b}_5 = 0 = \mathfrak{b}_4$,
  \item ${\bf 10 \, \ov 5_m \, \ov 5_H }$ is localized at $D_6$ point $\mathfrak{b}_5 = 0 = \mathfrak{b}_3$. 
\end{itemize}
Indeed $\mathfrak{b}_4 = 0 = \mathfrak{b}_5$  and $\mathfrak{b}_5 = 0 = \mathfrak{b}_3$ correspond to a single and double zero of $P_5$ in agreement with the order of ${\bf \ov 5}$ representations appearing in the coupling.
\begin{itemize}
  \item At $P=0=R$ but $(\mathfrak{b}_4,\mathfrak{b}_5) \neq (0,0)$ the singularity type enhances to $A_6$. 
\end{itemize}
These points are special in that they do not arise from the collision of two matter curves. They realize the coupling ${\bf  5_H \, \ov 5_m \, 1}$. See \tableref{tab:GaugeEnhanc} for a compact listing of all the aforementioned gauge enhancements.

Having introduced the geometrical properties which are necessary to realize an $SU(5)$ GUT we can now answer first questions about candidate del Pezzo geometries on which the GUT branes can be placed. We note that demanding a $\bf{10}$ matter curve poses strong restrictions on the genericity of the del Pezzo surface. This restriction is simplest discussed in the weak coupling picture where one obtains the $\bf{10}$ matter curve by intersecting the $SU(5)$ GUT seven-brane with its orientifold image in a Calabi-Yau orientifold \cite{Blumenhagen:2008zz}. The F-theory base $B$ is in this case the quotient of the Calabi-Yau manifold by the geometric orientifold involution, and brane and image-brane pairs will be identified in $B$ as in refs.~\cite{Blumenhagen:2009up, Collinucci:2009uh}. If the GUT brane is demanded to wrap a del Pezzo surface, one thus has to have two intersecting del Pezzo surfaces in the covering Calabi-Yau space. However, two intersecting del Pezzos cannot be generic and cannot be shrunk simultaneously to a point, due to the presence of the intersection curve. In fact, one can only shrink these del Pezzos to the intersection curve itself. In an F-theory base $B$ the GUT del Pezzo is thus also a non-generic del Pezzo surface which can only be shrunk to a curve. These are precisely the geometric features we have advertised in \secref{delPezzo}. Moreover, this gives a simple geometrical explanation for the no-go result of ref.~\cite{Donagi:2009ra}, where it was also suggested to weaken shrinkability to curves.
This, however, will enlarge the class of viable surfaces for local GUT models considerably. For example, a $dP_9$ surface, which is an elliptic fibration over $\bbP^1$, is shrinkable to a curve and still allows for a generalized decoupling limit.  GUT model building on $dP_9$ surfaces was considered within Type IIB orientifolds in \cite{Blumenhagen:2008zz}.

\subsection{Spectral cover construction}\seclabel{spectral_cover}
As we have reviewed, the geometric structure of the elliptic fibration $Y$ is specified by the sections $a_i$, and the appearance of singularities on the base $B$ can be understood in terms of their vanishing locus. A particularly efficient way to keep track of this information is via the so-called spectral cover construction. Even more importantly, this provides an elegant way to describe the gauge flux on the seven-branes in a consistent manner.

Spectral covers were first introduced in \cite{Friedman:1997yq, donagi-1997-1} in the context of heterotic compactifications on elliptically fibered threefolds, where they encode the information of the vector bundles embedded into $E_8 \times E_8$. By heterotic/F-theory duality it is clear that they have a natural appearance also for F-theory compactifications with a heterotic dual. More recently, however, it has been appreciated \cite{Hayashi:2009ge, Donagi:2009ra} that spectral covers are the natural language in which the geometry and gauge flux of F-theory compactifications even without (simple) heterotic duals is to be phrased. Recall that in models with a heterotic dual, the elliptic fourfold $Y$ also has the structure of a K3 fibration $\mathrm{K3} \surjto B_2$ over a complex surface $B_2$; i.e. the base space $B$ of the elliptic fibration $Y$ is itself  $\mathbb P^1$ fibered over $B_2$, which is the common basis also of the elliptically fibered threefold $Z$ on the heterotic side. In particular, $B_2$ is part of the singular component in the discriminant and would play the role of the GUT divisor $S$. The GUT divisor is therefore the base of a \emph{globally} defined fibration in models with heterotic dual. 

In general F-theory models, this is not the case. However, one can \emph{locally} view $S$ as the basis of a ALE fibration which describes the singularity structure along $S$ \cite{Beasley:2008dc}. For the Weierstrass models of $E_8$ type which we are considering here, the ALE fiber contains a distinguished set of two-cycles $\omega^I$ whose intersection form equals the Cartan matrix of $E_8$. For generic non-zero size of these two-cycles the $E_8$ symmetry is broken. If the divisor $S$ exhibits enhanced gauge symmetry $H$ this is because some of the $\omega^I$, called $\omega_G^I$ in the sequel, in the fiber shrink to zero size. The intersection matrix of the two-cycles $\omega_H^I$ with non-zero size is the Cartan matrix of the commutant $G \subset E_8$ of $H$.
 
This picture is of course very reminiscent of the breaking of $E_8$ to, say, $H$ in the heterotic string by means of a gauge bundle with structure group $G$. In F-theory, some of the degrees of freedom of the heterotic vector bundle are encoded purely geometrically, while others map to gauge flux. The geometric part is interpreted in the local field theory of \cite{Beasley:2008dc} as encoding the vacuum expectation value of the  Higgs field $\varphi$ associated with the normal fluctuations of the seven-brane. The spectral cover now is designed to describe the size of the non-zero two-cycles responsible for the breaking of $E_8$ to $H$ along $S$.

\subsubsection*{Spectral covers for an \boldmath$SU(5)$ model} 
In what follows we restrict ourselves to the spectral cover description of an $H=SU(5)_{\GUT}$ singularity along a divisor $S \subset B$ of the type introduced above. The complement of $SU(5)_{\GUT}$ in $E_8$ is denoted by $G= SU(5)_{\perp}.$ For  more  background and details on more general configurations we refer to \cite{Hayashi:2009ge, Donagi:2009ra}. 

The starting point is to construct an auxiliary non-Calabi-Yau threefold $X$ as a fibration over $S$ which encodes the singular geometry of $S$ in $B$. We will therefore think of $S$ either as a divisor on $B$ or as the base of a fictitious threefold $X$. The definition of $X$ is as the projectivized bundle over the GUT divisor $S$
\beq \label{defX}
  X = {\mathbb P} ({\cal O}_{S} \oplus K_S) ,\qquad p_X: X \surjto S .
\eeq
where $p_X$ is the projection to the base of the bundle. The base $S$ is viewed as the vanishing locus of the section $\sigma$ in $X$. This section satisfies the important relation
\beq \label{sigma2}
  \sigma \cdot \sigma = - \sigma\,  c_1(S).
\eeq
The manifold $X$ is not Calabi-Yau and has first Chern class
\beq
  c_1(X) = 2 \sigma + 2 c_1(S).
\eeq

In addition to the coordinates on the base $S$ of $X$ one introduces the projective coordinates $M,N$ to parametrize the fiber directions. These are defined such that on each fiber of $X$ they restrict to sections of ${\cal O}(1) \otimes K_S$ and of ${\cal O}(1)$, respectively. The spectral cover ${\cal C}^{(5)}$ is now defined as the divisor in $X$
\beq \label{SCCa}
  b_0 M^5 + b_2 M^3 N^2 + b_3 M^2 N^3 + b_4 M N^4 + b_5 N^5 = 0,
\eeq
with the $b_i$ corresponding to the ${\mathfrak b}_i$ appearing in the Tate equation \eqref{TateSU5b}. This time, however, they are interpreted as sections on $S$ pulled back to $X$. With the help of the adjunction formula $K_B|_S = K_S \otimes N_{S/B}^{-1}$ for hypersurfaces $S\subset B$ the $b_j$ can be expressed as  
\beq
  b_{6-j} \in H^0\big(S; K_B^{-j} |_S\otimes N_{S/B}^{1-j}\big).
\eeq
To keep with the standard notation of the spectral cover construction and for comparison with the dual heterotic constructions, one introduces the elements of $H^2(S; \mathbb Z)$
\beq
  \eta = 6 c_1(S) -t,  \qquad -t = c_1(N_{S/B}),
\eeq
and the coordinate $w$ is interpreted as a section of the normal bundle $N_{S/B}$. 
Then the sections $b_j$ are identified as
\beq \label{bisection}
  b_j \in H^0\big(S; {\cal O}(\eta - j c_1(S))\big) = H^0\big(S; {\cal O}((6 - j )c_1(S)  - t )\big).
\eeq
Equation \eqref{SCCa} is the projectivization of 
\beq \label{C5b}
  b_0 s^5 + b_2 s^3 + b_3 s^2 + b_4 s + b_5 = 0,
\eeq
viewed as a divisor of the total space of $K_S$. Here we use the variable $s=0$ to denote $S$ as the base of the total bundle $K_S$. In the sequel we will always think of the spectral surface ${\cal C}^{(5)}$ as a divisor of $X$. It is a 5-fold cover of the GUT brane $S$, to which we associate the projection
\beq
  \pi_5: {\cal C}^{(5)} \surjto S.
\eeq
Since $\sigma$ is the class of $s=0$ in $X$ and with the assignment \eqref{bisection} for $b_i$, we have cohomologically in $X$
\beq
  [{{\cal C}^{(5)}}] =5 \sigma + \pi_5^* \eta.
\eeq

Given the implicit underlying $E_8$ structure of the ALE fibration, the massless matter representations of $H= SU(5)$ can be understood as the irreducible representations $R_x$ in the decomposition ${\bf 248} \rightarrow \sum_x (R_x, U_x)$, 
\bea \label{matter-decomp-SU(5)}
  {\bf 248} \mapsto ({\bf 24},1) + (1,{\bf 24}) + [ ({\bf 10}, {\bf 5}) + ({\bf \ov 5}, {\bf 10}) + h.c.].
\eea

The matter curves of \tableref{tab:GaugeEnhanc} are understood as curves on $S$. The matter curve $P_{10}$ is the locus $b_5=0$ on $S$. It is associated with the spectral cover in the fundamental representation of $G= SU(5)_{\perp}$ because the ${\bf 10}$ appears as $ ({\bf 10}, {\bf 5})$ in \eqref{matter-decomp-SU(5)}. Let us also define the object ${\cal P}_{10}$ viewed as   a curve in $X$,
\beq
  {\cal P}_{10} =  {\cal C}^{(5)} \cap \sigma \subset X.
\eeq
Then the matter curve on $S$ is related to ${\cal P}_{10}$ as
\beq
  [P_{10}] = [{\cal P}_{10} ] |_{\sigma} =  (5 \sigma +\pi_5^*  \eta)|_{\sigma} = \eta - 5 c_1(S)
\eeq
with the help of \eqref{sigma2}, i.e.~the restriction of ${\cal P}_{10}$ is cohomologically equivalent to the matter curve $P_{10}$. 

The matter curve for the ${\bf \ov  5}$ on $X$ is more complicated and was analyzed in detail in the context of the heterotic string in \cite{Donagi:2004ia, Blumenhagen:2006wj, Hayashi:2008ba}. In view of the decomposition \eqref{matter-decomp-SU(5)} it should be regarded as the intersection 
\beq
 P_5= c_{\bigwedge^2 V} = \sigma \cap {\cal C}_{\bigwedge^2 V}
\eeq
of the spectral cover $\smash{{\cal C}_{\bigwedge^2 V}}$ which is associated with the antisymmetric representation of $SU(5)$ and hence provides a 10-fold cover of the GUT brane. For simplicity we will not introduce this spectral cover explicitly here, but note that, unlike ${\cal C}^{(5)}$, it is singular along a curve of double point singularities \cite{Donagi:2008ca, Hayashi:2008ba}. This curve of double points meets $P_5$ on $S$ in singular points. One therefore has to define the normalization $\smash{\ov c_{\bigwedge^2 V}}$ of $P_5$ as the actual, non-singular locus of the ${\bf \ov 5}$ matter viewed as a curve in $X$. What is easy to describe is the (branched) double cover ${\cal P}_5$ of this normalization. It arises as \cite{Donagi:2004ia, Blumenhagen:2006wj, Hayashi:2008ba}
\beq \label{P5a}
  {\cal P}_5 = \tau {\cal C}^{(5)} \cap  {\cal C}^{(5)} - {\cal C}^{(5)} \cap  \sigma -  {\cal C}^{(5)} \cap \sigma_t \equiv {\cal C}^{(5)} \cap ( \tau {\cal C}^{(5)} - \sigma_{\tau})   ,
\eeq
where $\tau$ is the ${\mathbb Z}_2$ involution $N \rightarrow -N$ acting on the spectral cover, $\sigma_\tau = \sigma + \sigma_t$, and  $\sigma_t = 3 (\sigma + \pi^*c_1(S))$ denotes the so-called tri-section. Then, 
\beq
  \ov c_{\bigwedge^2 V} = {\cal P}_5 /  \tau.
\eeq
This double cover is ramified at
\bea
R= {\cal C}^{(5)} \cdot ({\cal C}^{(5)} - \sigma_{\tau}) \cdot \sigma_{\tau} 
\eea
points. For concrete computations of the chiral matter we will only need the class $ [{\cal P}_5 ]$ in $X$. It can be easily be evaluated from \eqref{P5a} as
\beq
  [{\cal P}_5 ]= [{\cal C}^{(5)}] \cap [\sigma + \pi_5^*(\eta - 3 c_1(S))] .
\eeq 
This concludes the dictionary of how to treat our $SU(5)$ GUT model in terms of a spectral cover.

\subsection[Gauge flux in generic $SU(5)$ GUT models]{Gauge flux in generic \boldmath$SU(5)$ GUT models}\seclabel{gauge_fluxSU(5)}
So far all that the spectral cover approach has done for us is to rewrite the geometric data in a seemingly more complicated manner. Its actual power, however, becomes apparent once one includes gauge flux into the compactification. Along each seven-brane one can consider non-zero vacuum expectation values of the gauge flux $F$. In perturbative type IIB language, this flux is described by a holomorphic vector bundle whose structure group is embedded into the gauge group on the D7-brane. It is a necessary ingredient to obtain chiral matter along the matter curves. A detailed discussion of gauge flux in this language was given e.g.~in \cite{Blumenhagen:2008zz}. In F-theory, gauge flux arises by reduction of the four-form flux $G$ along the singular locus of the seven-branes with two legs parallel to the seven-brane~\cite{Denef:2008wq}. 

In the context of $SU(5)$ GUT theories one distinguishes between the gauge flux along the GUT divisor $S$ itself and the flux along the matter branes which constitute the complementary $\mathrm{I}_1$ locus of the discriminant. The first type of gauge flux takes values in $H=SU(5)_{\GUT}$  and will thus further break the $SU(5)$ GUT symmetry; it is the subject of \secref{hyper}. Gauge flux on the matter branes is given, on the Calabi-Yau fourfold $Y$, by gauge flux along the $\mathrm{I}_1$ component of the discriminant. In the spectral cover approach one makes use of the fact that ${\cal C}^{(5)}$ describes the geometry in the vicinity of $S$ and in particular the \emph{local} geometry of the $\mathrm{I}_1$ component.  

We think of the $SU(5)$ GUT symmetry as the effect of breaking the original $E_8$ symmetry  by giving a non-zero VEV to the two-cycles $\omega^I$ in the ALE fibration over $S$. Locally the flux on the $\mathrm{I}_1$ component can be described as a VEV for the field strength on $S$ with values in the complementary $G= SU(5)_{\perp}$. In the machinery of the spectral cover, this flux is given in terms of the so-called spectral line bundle ${\cal N}$ along ${\cal C}^{(5)}$ defined by its first Chern class
\beq
  c_1({\cal N}) \in H^{(1,1)}({\cal C}^{(5)}; \mathbb Z) .
\eeq  
Since ${\cal C}^{(5)}$ is a five-fold cover of $S$ in $X$ one can push this line bundle forward to $S$ via ${\pi_5}_\ast$. This defines a rank-5 vector bundle 
\beq
  V = \pi_{5 \ast} {\cal N} 
\eeq
on $S$. Following the general logic of the ALE fibration over $S$ the structure group of this bundle $V$ is identified with the commutant $SU(5)_{\perp} \subset E_8$ of the GUT $SU(5)$ along $S$ and therefore indeed associated with the flux on $\mathrm{I}_1$.

In view of later applications, let us be a little bit more general and consider a general rank $n$ bundle $V$ defined in this manner. For the first Chern class $c_1({\cal N}) \in H^2({\cal C}^{(5)}; \mathbb Z)$ of the spectral bundles one can make the general decomposition ansatz \cite{Friedman:1997yq, Andreas:2004ja}
\beq \label{c1N}
  c_1(\mathcal{N}) = \frac{r}{2} + \gamma.     
\eeq
Here we have abbreviated
\beq
  r =  {-\,c_1({{\cal C}^{(n)}})+ \pi_n^{\ast} c_1(S)} , \qquad 
  \gamma =  \frac{1}{n}\,  \pi_n^{\ast} c_1({ V})  + \gamma_u  ,
\eeq
where $r$ denotes the ramification divisor of the $n$-fold cover ${\cal C}^{(n)}$ and $\gamma_u$ is chosen such that it satisfies $\pi_{n \ast} \gamma_u =0$. This yields 
\beq \label{gammau}
  \gamma_u=\lambda\, (n\sigma-\pi_n^{\ast}\eta +{n}\pi_n^{\ast}c_1(S)) ,
\eeq
for a number $\lambda \in \mathbb{Q}$. Let us further parametrize $c_1({ V})$ by some element  $\zeta \in H^2(S; {\mathbb Z})$ \cite{Andreas:2004ja},
\beq
  \zeta = c_1({V}) ,
\eeq
which allows for direct comparison with the formulas for heterotic models. The parameter $\lambda \in \mathbb Q$ is subject to certain constraints to be discussed shortly. Note that this bundle exists for generic complex structure since it only involves  $\sigma$ and the pullback of classes from $S$. For the time being we are interested in an $SU(5)_{\perp}$ bundle $V$, corresponding to $n=5$ and $\zeta = 0$.

The parameter $\lambda \in \mathbb Q$ has to be chosen such that $c_1({\cal N})$ defines an integer class in $H^2({\cal C}^{(n)}; \mathbb Z)$. On the non-Calabi-Yau space $X$ the adjunction formula leads to
\beq \label{c1Cn}
  - c_1({{\cal C}^{(n)}}) = (n-2) \sigma + \pi_n^* ( \eta -2 c_1(S)).
\eeq
Putting everything together, we have 
\beq \label{linebundle}
  \bal
    c_1(\mathcal{N}) = {} & {} - \sigma + n\left({\textstyle  \frac{1}{2}+\lambda }\right)\,\sigma + \left({\textstyle \frac{1}{2} - \lambda}\right) \pi_n^\ast \eta \\
    & {} + \left( {\textstyle - \frac{1}{2}+ n \lambda} \right) \pi_n^\ast c_1(S) + {\textstyle \frac{1}{n}}\, \pi_n^\ast \zeta.
  \eal
\eeq
For an $SU(5)$ bundle, integrality of $c_1({\cal N})$ therefore puts the value of $\lambda \in {\mathbb Q}$ subject to the constraints 

\beq \label{Ninteger}
    5\left ({\textstyle \frac{1}{2}+  \,\lambda }\right)  \in {\mathbb Z}\ ,\qquad 
    \left({\textstyle \frac{1}{2}-\lambda}\right)\, \eta +\left({\textstyle 5 \lambda - \frac{1}{2}}\right) c_1(S) \in H^2(S; {\mathbb Z})\ .
\eeq

The spectral cover also allows for a local reconstruction of the four-form flux $G$.  To this end one exploits the fact \cite{Donagi:2008ca, Donagi:2009ra} that a two-form on ${\cal C}^{(5)}$ defines a two-form on $S \subset X$ with values in $H^{0}(f)$, where $f$ is the fiber of the cover $\pi_5: {\cal C}^{(5)} \surjto S$. $H^{0}(f)$ is generated by  the zero-forms $\omega_0^I$ associated with the five points on the spectral cover over $S$. In the local ALE fibration each of these points is promoted to one of the five non-zero two-cycles $\omega^I$. Therefore the two-form $\gamma_u \in H^{(1,1)}({\cal C}^{(5)}; \mathbb Z)$ is associated with a two-form on $S$ with values in a basis for these $\omega^I$. This was formalized in \cite{Donagi:2008ca, Donagi:2009ra} by writing
\beq \label{FG}
  \gamma_u = F_I \wedge \omega_0^I \quad \longrightarrow \quad G- q S^* = F_I \wedge \omega^I.
\eeq
Note that one has to subtract a suitable multiple of the Poincar\'e dual $S^*$ to ensure D-flatness \cite{Donagi:2009ra}.

\subsubsection*{Induced 3-brane tadpole}
Given this relation between the gauge flux and the form $\gamma$ in F-theory, it is furthermore natural to conclude \cite{Curio:1998bva}
\bea
\label{Gflux}
\frac{1}{2} \int_Y G \wedge G = - \frac{1}{2} \int_S \pi_{n*}(\gamma^2).
\eea
The identification \eqref{Gflux} also allows one to compute the flux induced 3-brane charge that enters \eqref{D3a} via the spectral cover bundle. For a single $G=SU(n)$ bundles the result is \cite{Curio:1998bva, Andreas:2004ja}
 \bea
 \label{piga}
 \pi_{n*}(\gamma^2) = - \lambda^2 n  \eta(\eta - n c_1(S)).
\eea

Similar to our remarks at the end of \secref{GeomTad} in connection with the computation of $\chi(Y)$ the following words of caution are in order: The explicit formula for $G$ is certainly correct locally, but more generally $G$ will depend also on the normal coordinates to $S$ because generically $Y$ is not a global fibration over $S$. Recall that in the computation of the Euler characteristic of the singular $Y$ we observed by explicit checks that this plays no role. We therefore extrapolate the success of this computation also to the flux sector, in the sense that we will assume that \eqref{Gflux} is valid beyond F-theory models with a heterotic dual.

\subsubsection*{Chiral Matter}
The primary reason for the inclusion of gauge flux on the matter branes is that this opens up the possibility of a chiral matter spectrum on the matter curves. The multiplicity of a  given representation $R_x$ under $SU(5)_{\GUT}$ is determined by the cohomology groups of the spectral cover bundle $V$ associated with the representation $U_x$ of $SU(5)_{\perp}$ under the decomposition ${\bf 248} \mapsto \sum_x (R_x, U_x)$.

In the heterotic context of the spectral cover bundle, the computation of the matter content was initiated in \cite{Friedman:1997yq, Donagi:2004ia}. In F-theory, it is more appropriate to think in terms of extension groups, which are known to count massless matter at the intersection of two branes \cite{Katz:2002gh}. In fact this IIB-/F-theory picture had been used in \cite{Blumenhagen:2006wj} as inspiration to determine the spectrum on the heterotic side. 


The exact spectrum in the ${\bf 10}$ arises in the sector of the spectral cover ${\cal C}^{(5)}$ with bundle ${\cal N}$ and the zero section $\sigma$ carrying the trivial line bundle. Then the relevant groups are \cite{Hayashi:2009ge, Donagi:2009ra} 
\beq \label{coho_10}
  {\rm Ext}^i(\iota_* {\cal O}_{\sigma};  { j}_* {\cal N} ) = H^{(i-1)} ({\cal C}^{(5)} \cap \sigma; {\cal N} \otimes K_S |_{{\cal C} \cap \sigma}), \quad i = 1,2,
\eeq
where $\iota$ and ${j}$ embed $\sigma$ and the spectral surface into $X$. The chiral index of the ${\bf 10}$ is therefore computed via the Hirzebruch-Riemann-Roch theorem as
\beq \label{chi10}
  \bal
    \chi_{\bf 10} &= \chi(\ccP_{10}, {\cal N} \otimes K_S |_{\ccP_{10}} ) = \int_{\ccP_{10}} c_1({\cal N}) + c_1(K_S) + \frac12 c_1(\ccP_{10}) |_{\ccP_{10}} \\ 
    &= \int_{\ccP_{10}} \gamma - \frac{1}{2}c_1({\cal C}^{(5)}) - \frac12 \pi_5^*(c_1(S)) + \frac12 c_1 (\ccP_{10}) |_{\ccP_{10}} = \int_{\ccP_{10}} \gamma.
  \eal
\eeq
The last equality follows by adjunction for the curve $\ccP_{10} = \sigma |_{{\cal C}^{(5)}}$  and with $\sigma|_{\sigma} = - \sigma c_1(S)$. Given the identification of the piece $\gamma$ in ${\cal N}$ on ${\cal C}^{(5)}$ with the flux on the matter branes, this reproduces the intuition from Type IIB/F-theory that the chirality is given by integrated flux over the matter curve. Recall that in eqn.~\eqref{chi10}, $\gamma$ is an element of $H^2({\cal C}^{(5)}; \mathbb Z)$ and $\ccP_{10}$ is interpreted as the curve $\sigma \cdot {\cal C}^{(5)}$ in $X$. For the $SU(5)$ bundle at hand, this can be further evaluated on $P_{10} \subset S$ as
\beq \label{chi10a}
  \chi_{\bf 10} =\sigma \cdot {\cal C}^{(5)} \cdot \gamma =  - \lambda  \int_S  \eta  \underbrace{( \eta - 5 c_1(S))}_{P_{10}}.
\eeq

In general there will be non-chiral pairs of matter invisible to the index. In favorable cases, however, \eqref{coho_10} is uniquely determined by the index because a negative degree line bundle on a smooth curve has no sections. 

Applying a similar logic, the massless modes in the ${\bf \ov 5}$ representation should be given by
\beq \label{coho_5}
  {\rm Ext}^i(\iota_* {\cal O}_{\sigma};  { j}_* {\cal N}_{\bigwedge^2 V} ) = H^{(i-1)} ({c}_{\bigwedge^2 V}; {\cal N}_{\bigwedge^2 V} \otimes K_S |_{ c_{\bigwedge^2 V}}), \quad i = 1,2.
\eeq
However, due to the singular structure of $\smash{c_{\bigwedge^2 V}}$ one must actually work on the normalization $\smash{\ov c_{\bigwedge^2 V} = {\cal P}_5/ \tau}$, and the relevant cohomology groups replacing the right-hand side of \eqref{coho_5} are  \cite{Blumenhagen:2006wj} (see also \cite{Hayashi:2008ba, Hayashi:2009ge})
\beq
  H^{(i-1)} (  {\cal P}_5/ \tau; L  \otimes K_X^{-\frac12}  |_{ {\cal P}_5 / \tau}  ), \qquad  c_1(L)|_{ {\cal P}_5 / \tau} =c_1 ( {\cal N} \otimes K_S^{\frac12} ) |_{{\cal P}_5} - \frac {R}{2},
\eeq
with $R = {\cal C}^{(5)} \cdot ({\cal C}^{(5)} - \sigma_{\tau}) \cdot \sigma_{\tau}$ the number of ramification points of the cover ${\cal P}_5 \rightarrow  {\cal P}_5 /\tau $.\footnote{Up to the factor $K_X^{-1/2}$ this is equ. (104) of \cite{Blumenhagen:2006wj} after fixing the typo $+R/2 \rightarrow - R/2$.} Note that we had to include a correction term $K_X^{-1/2}$, which is trivially absent in the context of heterotic computations on a Calabi-Yau. If one includes this correction, the chiral index can be determined entirely on ${\cal P}_5$ as
\beq
  \bal
    \chi_{\bf \ov 5} &= \int_{\ccP_5} \left( c_1({\cal N}) + \frac12 c_1(K_S) + \frac14 c_1(X) \right) + \frac12 c_1(  {\cal P}_5 / \tau   ) |_{{\cal P}_5 / \tau } - \frac{R}{2}      \nonumber  \\   
    &=  \int_{\ccP_5} \gamma = [\ccP_5] \cdot \gamma.
  \eal
\eeq
For further details of this computation we refer to \cite{Blumenhagen:2006wj}. In fact it is easy to convince oneself that $\chi_{\bf 10} = \chi_{\bf \ov 5}$ as required for a consistent bundle by anomaly cancellation. 

The above considerations are  special cases of the following more general situation, which we will need for later applications: Consider two spectral covers ${ C}_1$ and ${C}_2$ and suppose the matter associated with the bifundamental representation $V_1 \otimes V_2$ of the corresponding bundles is localized on the curve
\beq \label{Pab}
  {\cal P} = \tau C_1 \cap C_2 - \tau C_1 \cap C_R 
\eeq
on $X$. The divisor $C_R$ has to be introduced in order to account for the ramification points. More precisely, the subtraction in \eqref{Pab} is required if the matter curve is singular over $R=  \tau C_1 \cdot (C_2 - C_R) \cdot  C_R $ points, as in the example of the antisymmetric representation. Then we propose that the massless matter is counted by the cohomology groups
\beq \label{coho_ab}
  H^{(i-1)} ( {\cal P}; {\cal N}_1 \otimes   {\cal N}_2 \otimes K_S  \otimes K_X^{-\frac12} \otimes {\cal O}(-R/2) |_{\mathcal{P}}), \quad i = 1,2.
\eeq
Indeed the associated index is correctly given by
\beq
  \chi = \int_{\mathcal{P}} (\gamma_1 + \gamma_2).
\eeq
For the proposal \eqref{coho_ab} to pass this consistency check we had to include two correction terms, $K_X^{-1/2}$ and  ${\cal O}(-R/2)$ compared to naive expectations. Strictly speaking, the factor $K_X^{-1/2}$  has to appear also in \eqref{chi10}, but there it vanishes trivially because $\sigma \cdot c_1(X) = \sigma \cdot (\sigma + c_1(S)) = 0$. The term ${\cal O}(-R/2)$ accounts for the ramification points on the matter curve. It would be interesting to derive this formula from first principles by an explicit evaluation of the spectral sequences for the underlying extension groups.

\subsection[{GUT breaking by hypercharge flux via $S\big[U(5)\times U(1)\big]$ bundles}]{GUT breaking by hypercharge flux via \boldmath$S\big[U(5)\times U(1)\big]$ bundles}\seclabel{hyper}
An important issue in the construction of $SU(5)$ GUT models is the breaking of $SU(5)$ to $SU(3) \times SU(2) \times U(1)_Y$. A simple mechanism to accomplish this task is by means of yet another type of gauge flux which transforms as $U(1)_Y \subset SU(5)_{\GUT}$. In the context of heterotic compactifications this method was applied systematically in \cite{Blumenhagen:2005ga, Blumenhagen:2006ux, Weigand:2006yj}. The $E_8$ symmetry is first broken to $SU(5)_{\GUT}$ by means of an $SU(5)_{\perp} $ bundle and then an extra line bundle with structure group $U(1)_Y$ is responsible for the breaking of the GUT symmetry. Concretely, this leads to the decomposition of the GUT spectrum
\beq \label{splitting} 
  \bal
    {\bf 24}        &\mapsto  ({\bf 8},{\bf 1})_{0_Y} + ({\bf 1},{\bf 3})_{0_Y} + ({\bf 1},{\bf 1})_{0_Y} + ({\bf 3},{\bf 2})_{5_Y}+  ({\bf \ov 3},{\bf 2})_{-5_Y},   \\ 
    \ov {\bf 5}\ \, &\mapsto  (\ov{\bf 3},{\bf 1})_{2_Y} + ({\bf 1},{\bf2 })_{-3_Y},   \\ 
    {\bf 10}        &\mapsto ({\bf 3},{\bf 2})_{1_Y} +  ({\bf \ov 3},{\bf 1})_{-4_Y} + ({\bf 1},{\bf 1})_{6_Y},   \\ 
    {\bf 5}_H       &\mapsto  ({\bf 3},{\bf 1})_{-2_Y} + ({\bf 1},{\bf2 })_{3_Y},\qquad  
    \ov {\bf 5}_H    \mapsto  (\ov {\bf 3},{\bf 1})_{2_Y} + ({\bf 1},{\bf2 })_{-3_Y}, 
  \eal
\eeq
which is the same as encountered in the Georgi-Glashow $SU(5)$ GUT model. The decomposition of the adjoint {\bf 24} provides the Standard Model gauge bosons along with some exotics. From $\mathbf{10\oplus\bar 5\oplus 1}$ an entire left-handed matter generation is obtained, whereas $\mathbf{5}_H$ and $\mathbf{\bar 5}_H$ give the Higgs.

In the present context the GUT group breaking $U(1)_Y$ flux is described by a  2-form $F_Y \in H^{(1,1)}(S)$ along the GUT brane \cite{Beasley:2008kw} with corresponding F-theory four-form flux 
\beq
  G_4  = F_Y \wedge \omega^I_H.
\eeq
Recall that the $\omega^I_H$ represent the zero-size two-cycles in the ALE fiber over $S$ which span the Cartan matrix of the hitherto unbroken $SU(5)_{\GUT}$ symmetry. This $F_Y$ is the curvature of the hyperflux line bundle $L_Y$  on $S$.

There are two challenges associated with this strategy. First the vectorlike states $({\bf 3},{\bf 2})_{5_Y}+  ({\bf \ov 3},{\bf 2})_{-5_Y}$ arising from the decomposition of the $ {\bf 24}$ in \eqref{splitting} have to be absent at the massless level for phenomenological reasons. Given the fact that these states are counted by $H^*(S; L_Y^{\pm 5})$, this condition is hard to achieve for $c_1 (L_Y) \subset H^2(S; \mathbb Z)$ as noted in \cite{Beasley:2008kw}. 
 
This problem can be circumvented by a certain twisting procedure of $L_Y$ which was first applied in \cite{Blumenhagen:2005ga, Blumenhagen:2006ux} in the heterotic context. Given the significance of the spectral cover construction also in the F-theory context, this mechanism can equally be put to work here. Group-theoretically one invokes a rank-five spectral cover bundle of non-trivial first Chern class $c_1(V) \neq 0$. Then the bundle
\beq
  \label{U5}
  V \oplus {\cal L}_Y, \qquad c_1(V) + c_1({\cal L}_Y) = 0
\eeq
is embedded as an $S[U(5) \times U(1)_Y]$ bundle into $E_8$. It is important that $c_1(V)$ and $c_1({\cal L}_Y)$ are correlated as in \eqref{U5}. The decomposition of the adjoint representation of $E_8$ reads
\beq \label{breaking}
  \bal
    E_8{\,\,} & \longrightarrow{\,\,\,} SU(5) \times SU(3)\times SU(2) \times U(1)_Y \\
    {\bf 248} & \mapsto 
    \left\{\begin{array}{l}
      ({\bf 24}; {\bf 1}, {\bf 1})_0 + ({\bf 1}; {\bf 1}, {\bf 1})_0 + ({\bf 1};{\bf 8}, \bf 1)_0 +({\bf 1}; {\bf 1},{\bf 3})_0 \\
      ({\bf 5}; {\bf 3},{\bf 2})_1 + ({\bf 1}; {\bf 3},{\bf 2})_{5} + c.c.\\
      ({\bf 10}; {\bf \ov 3},{\bf 1})_2 + ({\bf 5}; {\bf \ov 3},{\bf 1})_{-4} + c.c.\\
      ({\bf 10}; {\bf 1},{\bf 2})_{-3} + ({\bf 5}; {\bf 1},{\bf 1})_{6} + c.c.
    \end{array}\right\}.
  \eal
\eeq

The Cartan generators of the structure group of  $V \oplus {\cal L}_Y$ are embedded as
\beq
(1,1,1,1,1,-5)
\eeq
into $SU(6) \subset E_8$. This effectively attributes $U(1)_Y$ charge $+1$ to the fundamental representation of $V$ and $-5$ to ${\cal L}_Y$. We summarize the spectrum in \tableref{spec_U5}.

\begin{table}[ht]
  \centering
  \renewcommand{\arraystretch}{1.3}
  \begin{tabular}{c|c|cl}
    $SU(3)\times SU(2)\times U(1)_Y$ & bundle                                  & \multicolumn{2}{c}{Standard Model particles}$_\big.$ \\  
    \hline \hline
    $({\bf 3},{\bf 2})_{1}$        & $V$$^\big.$                               & $q_L$              & L-handed quark\\
    $({\bf 3},{\bf 2})_{{5}}$      & ${\cal L}_Y^{-1}$$_\big.$                 & $-$                & (exotic matter) \\
    \hline
    $({\bf \ov 3},{\bf 1})_{2}$    & $\bigwedge^2 V$$^\big.$                   & $\bar d_L = d^c_R$ & L-handed down antiquark \\
    $({\bf \ov 3},{\bf 1})_{-{4}}$ & $V\otimes {\cal L}_Y$$_\big.$             & $\bar u_L = u^c_R$ & L-handed up antiquark \\
    \hline
    $({\bf 1},{\bf 2})_{-3}$       & $\bigwedge^2 V\otimes {\cal L}_Y$$^\big.$ & $l_L$              & L-handed lepton \\
    $({\bf 1},{\bf 1})_{6}$        & $V\otimes {\cal L}_Y^{-1}$ $_\big.$       & $\bar e_L = e^c_R$ & L-handed antielectron \\
  \end{tabular}
  \caption{\small Dictionary between the standard model representations and bundles for the direct breaking $E_8 \longrightarrow SU(5)\times\big[ SU(3)\times SU(2)\times U(1)_Y \big]$.}
  \tablelabel{spec_U5}
\end{table}

Note in particular that now the $({\bf 3},{\bf 2})_{5}$ is associated only with one power of ${\cal L}_Y$ and thus easier to avoid. This effectively realizes the suggestion of \cite{Beasley:2008kw} to use suitably fractional line bundles on the branes.

A second challenge is that the $U(1)_Y$ gauge boson generically acquires a mass via Chern-Simons couplings to the closed string background fields. E.g.~in the heterotic case, the relevant couplings involve the Neveu-Schwarz B-field. In \cite{Blumenhagen:2006ux} this problem was solved by realizing $U(1)_Y$ as the massless linear combination of two massive $U(1)$ symmetries. This is, however, at the cost of sacrificing straightforward gauge coupling unification, which can only be re-installed in the strong coupling limit of heterotic M-theory \cite{Tatar:2008zj}. In F-theory/Type IIB, the complication of a massive $U(1)_Y$ can be circumvented in an elegant manner if $F_Y$ is an element of the relative cohomology of $S \subset Y$. This means that the Poincar\'e dual two-cycles, while non-trivial elements of $H_2(S)$, are the boundary of a 3-chain in $Y$ \cite{Beasley:2008kw}. As we will recall in section \ref{sec_Pheno}, however, problems with straightforward gauge coupling unification arise in F-theory with hyperflux as well.

The twisting also affects the computation of the flux contribution to the 3-brane tadpole in \eqref{D3a}. As discussed around \eqref{piga} the general procedure is to compute  $ \pi_{n*}(\gamma^2)$, but now both the $U(5)$ bundle ${\cal V}$ and the line bundle ${\cal L}_Y$ contribute. The exact appearance of the respective $\pi_{n*}(\gamma^2)$ in $N_{3}$ is governed by the group theoretic embedding. Again a look at the heterotic side of the medal is helpful: There, for compactifications with $U(n)$ bundles, the terms $-c_2(V_i)$ in \eqref{M5} of \appref{Tadpoles_sec} are representative for the more general $\frac{1}{4 (2 \pi)^2}     {\rm tr}(\ov F^2) $  with \cite{Blumenhagen:2005ga, Blumenhagen:2006ux}
\beq \label{traces W_1}
  {\rm tr}(\ov F^2)= \frac{1}{30}\, {\rm Tr}(\ov F^2) = \frac{1}{30} \, \sum_{x } 2 (2 \pi)^2 \left( {\rm ch}_2(U_x) \times {\rm dim} (R_x) \right).
\eeq  
Here ${\rm tr}$ and ${\rm Tr}$ denote the trace over the fundamental and adjoint representation of $E_8$, respectively. The sum is over all irreducible representations $(U_x, R_x)$ under the decomposition of $E_8$ into $(G,H)$ for a bundle of structure group $G$. For the twisted $S[U(5) \times U(1)_Y]$ embedding in \tableref{spec_U5} one evaluates
\beq \label{tracesSU5}
\frac{1}{4 (2 \pi)^2}   \,  {\rm tr}(\ov F^2)=    {\rm ch}_2({\cal V}) +  {\rm ch}_2({\cal L}) .
\eeq  
Furthermore, for a $U(n)$ bundle equ. \eqref{piga} generalizes to 
\beq \label{pigb}
  \pi_{n*}(\gamma^2) = - \lambda^2 n  \eta(\eta - n c_1(S)) + \frac{1}{n} \zeta^2.
\eeq
Therefore 
\beq
  -\frac12 G^2=  \frac{1}{2}\left( - 5 \lambda^2   \eta( \eta - 5 c_1(S)) + \left(\frac{1}{5} +1\right) \zeta^2  \right).
\eeq
The complete 3-brane tadpole equation is thus
\beq
  \bal
    N_3 = {}& \frac{\chi^*(Y)}{24}  - 305 \int_S c^2_1(S) - 15 \int_S  (  \eta^2 - 9 \eta c_1(S) )  \\ 
    &{}+ \left(\frac58 - \frac52 \lambda^2\right) \int_S \eta (\eta - 5 c_1(S)) + \frac35 \int_S \zeta^2,
  \eal
\eeq
where $\chi^*(Y)$ is the Euler characteristic as evaluated by using the base $B$ in \eqref{smooth_chi}.

\section{\boldmath$S\big[U(4)\times U(1)_X\big]$ split spectral covers}\seclabel{sec_split}

In this section we present a refined spectral cover construction which ensures the absence of dimension 4 proton decay operators. In \secref{split_covers} we introduce the appropriate split spectral cover and analyze the matter curves and Yukawa couplings. The split spectral cover bundle and its quantization conditions are discussed in \secref{sec_Bundle}. This split spectral cover also leads to modified chiral indices for the matter representation. The GUT breaking by a suitably twisted hypercharge bundle and the induced 3-brane tadpole are analyzed in \secref{split_GUTbreaking}. Finally, we comment on the D-term supersymmetry conditions in \secref{SUSY_sec}.

\subsection{Split spectral covers}\seclabel{split_covers}
In the generic spectral cover  construction of ${\cal C}^{(5)}$ both the ${\bf \ov 5_m}$ and the ${\bf 5_H + \ov 5_H}$ are localized on the same single curve $P_{5}$ of $A_5$ enhancement. This is phenomenological disfavored since along  with the ${\bf 10 \, 10\,  5_H }$ and  ${\bf 10 \, \ov 5_m  \, \ov 5_H }$ also the dangerous operator ${\bf 10 \, \ov 5_m  \, \ov 5_m }$ leading to dimension 4 proton decay is generated. A necessary condition to avoid the latter is that the  ${\bf \ov 5_m}$ and the ${\bf 5_H + \ov 5_H}$ live on distinct curves \cite{Beasley:2008kw}. It was  realized in \cite{Tatar:2009jk} that this factorization of the 5-curve is not sufficient, but rather that the whole spectral cover has to split into two disjoint components. Effectively this means that $SU(5)_{\perp}$ is replaced by $SU(4) \times U(1)_X$. All matter is now charged under the massive $U(1)_X$ factor, which only allows the desired Yukawa couplings while forbidding the dangerous operators \cite{Tatar:2006dc}.

Let us therefore specify to the situation of a factorized divisor
\beq
  {\cal C}^{(5)} = {\cal C}^{(4)} + {\cal C}^{(1)} ,
\eeq
each endowed with projections $\pi_4$ and $\pi_1$, respectively.\footnote{For ease of notation we  will oftentimes not indicate the subscripts in the projections $\pi_4$ and $\pi_1$.} The details of this factorization were worked out in \cite{Marsano:2009gv}. The split corresponds to the factorization of \eqref{C5b} into
\beq \label{C4-1}
  (c_0 s^4 + c_1 s^3 + c_2 s^2 + c_3 s + c_4) (d_0 s + d_1) = 0.
\eeq

Generically, there arise 2 matter curves for the ${\bf 10}$ representation at the intersection of the spectral cover with the GUT brane $S: s=0$, i.e. at
\beq \label{10curveb}
  c_4 = b_5 = 0, \qquad d_1 = 0.
\eeq
To avoid the second type of ${\bf 10}$ one sets $d_1 = const. \neq 0$ (so $d_1=1$ without loss) by defining it as an element of $H^0(S; {\cal O}_S)$. Comparison of \eqref{C4-1} with \eqref{C5b} allows us to express the sections $b_i$ as  \cite{Marsano:2009gv}
\beq \label{bcrel}
  b_5 = c_4, \quad b_4 = c_3 + c_4 d_0, \quad b_3 = c_2 + c_3 d_0, \quad b_2 = c_1 + c_2 d_0, \quad b_0 =  c_0 d_0
\eeq
subject to the constraint
\beq \label{cconstr}
  c_0 = -c_1 d_0.
\eeq
This identifies the coefficients appearing in the factorized polynomials as sections
\beq \label{split_sections}
  \bal
    d_1 &\in H^0(X; {\cal O}), \qquad d_0 \in H^0(X; p_X^* (TS)), \\
    c_n &\in H^0(X; {\cal O}(p_X^*(\eta - (1+n) c_1(S))). 
  \eal
\eeq

In particular the class of ${\cal C}^{(5)}$ splits into
\beq
  [{{\cal C}^{(4)}}] = 4 \sigma + \pi_4^* \tilde \eta, \qquad [{{\cal C}^{(1)}}] =  \sigma + \pi_1^* c_1(S),
\eeq
where
\beq
  \tilde \eta = \eta - c_1(S), \qquad \eta = 6 c_1(S) + c_1(N_{S}).
\eeq
Note that in concrete examples it has to be ensured that the factorization \eqref{cconstr} is indeed possible.

The factorization of the spectral cover, or equivalently the choice of sections as in \eqref{bcrel} for the Weierstrass model, has important consequences for the matter curves of the GUT divisor. The ${\bf 10}$ matter curve is realized as the locus $c_4 =0$ of cohomology class
\beq
  [\ccP_{10}]|_{\sigma} =  \eta - 5 c_1(S) = \tilde \eta - n c_1(S), \quad n=4.
\eeq
The matter curve \eqref{curve5a} for the ${\bf 5}$ factorizes as 
\beq
  \underbrace{(c_3(c_2 + c_3 d_0) - c_1 c_4)}_{P_H}  \underbrace{(c_2 + d_0(c_3 + c_4 d_0))}_{P_m} = 0
\eeq
so that the ${\bf 5}_H + {\bf \ov 5}_H$ and the matter ${\bf \ov 5}_m$ are localized on two different curves.

These two matter curves intersect at two distinct sets of points, described respectively by
\beq \label{PmPH}
  \bal
    &(\mathrm{I}) && c_4 = 0 \ \cap \ c_2 + c_3 d_0 = 0,  \\
    &(\mathrm{II}) && c_1 + c_3 d_0^2 = 0 \ \cap \ c_2 + c_3 d_0 + c_4 d_0^2 = 0, \\
    &&&{\rm but} \quad c_4 \neq 0,  c_2 + c_3 d_0 \neq 0.
  \eal
\eeq
on $S$. The first type of intersection points  is just the codimension-3 locus of $D_6$ enhancement associated with the ${\bf 10 \, \ov5_H \, \ov5_m}$ Yukawa couplings. At the second set of points, also the polynomial $R$ in \eqref{Delta_SU5} vanishes. This identifies them as the points of $SU(7)$ enhancement, interpreted as the locus of ${\bf 5_H \, \ov 5_m \, 1}$. Altogether the Yukawas are thus realized on 
\beq
  \bal
    & D_6: && c_4=0 \ \cap \ c_2 + c_3 d_0 = 0, \\
    & E_6: && c_4=0 \ \cap \  c_3 + c_4 d_0=0,  \\
    & A_6: && c_1 + c_3 d_0^2 = 0 \ \cap \ c_2 + c_3 d_0 + c_4 d_0^2 = 0, \\
    &&& {\rm but} \quad c_4 \neq 0,  c_2 + c_3 d_0 \neq 0.
  \eal
\eeq

Finally, we will also need at least the cohomological expressions for $[\ccP_H]$ and $[\ccP_m]$ as curves in $X$. These can be understood as a decomposition of \eqref{P5a} under the split ${\cal C}^{(5)} \rightarrow {\cal C}^{(4)} \cup {\cal C}^{(1)}$. As was worked out in great detail in \cite{Marsano:2009gv}, $\ccP_H$ and $\ccP_m$ derive from the respective pieces $\tau {\cal C}^{(4)} \cap  {\cal C}^{(4)} $ and $\tau {\cal C}^{(1)} \cap {\cal C}^{(4)}$ after appropriate subtractions as in \eqref{P5a}. Cohomologically this leads to \cite{Marsano:2009gv}  
\beq \label{classPH}
  \bal
    {}[\ccP_H] &=  2 \sigma \cdot \pi^*(2 \tilde\eta - 5 c_1(S)) + \pi^*(\tilde\eta - c_1(S)) \cdot \pi^*(\tilde\eta - 2 c_1(S)),  \\
    {}[\ccP_m] &=    \sigma \cdot \pi^*(  \tilde\eta - 2 c_1(S)) + \pi^*(c_1(S)) \cdot \pi^*(\tilde\eta - 2 c_1(S)).
  \eal
\eeq

As for the unfactorized spectral cover, the object $[\ccP_H] $ is the double cover of the normalization of the singular Higgs curve $P_H$ of class $[P_H] = 2 \tilde \eta  -5 c_1(S)$ associated with the spectral cover for ${\cal C}_{\bigwedge^2 V}$. The subtracted term accounts for the ramification points of this double cover. For later purposes we note that 
\bea \label{Pnu}
  [\tau {\cal C}^{(1)} \cap {\cal C}^{(4)}] =   [\ccP_m]  +  [\ccP_{\nu}], \qquad   [\ccP_{\nu}] = 2 (\sigma + \pi^* c_1(S) ) \cdot \pi^* c_1(S).
\eea
The so-defined curve $ \ccP_{\nu}$ is thus the component of the  intersection of ${\cal C}^{(4)}$ and ${\cal C}^{(1)}$ away from $S$ as is evident from $\sigma \cdot (\sigma + \pi^* c_1(S) )  =0$, see equ. (\ref {sigma2}).

\subsection{Split spectral cover bundle}\seclabel{sec_Bundle}
The split of the spectral cover implies a corresponding factorization of the structure group of the rank-5 bundle obtained from ${\cal C}^{(4)} + {\cal C}^{(1)}$ from $SU(5)_{\perp}$ to $S[ U(4) \times U(1)_X]$. From \secref{hyper} we are familiar with this $S[U(N) \times U(1)]$ construction, even though there the $U(1)$ part lives on $S$ and not on the spectral cover. The group theoretic treatment is very similar, though. 

At the level of the spectral cover the correlation of the $U(4)$ and the $U(1)$ part becomes effective through the constraint \eqref{cconstr}, which is necessary for a consistent embedding of the two groups into $SU(5)_{\perp} \subset E_8$. The gauge flux on the matter branes is encoded in an $S[U(4) \times U(1)_X]$ bundle ${W}$ on $S$ given by
\beq
  {W} = V \oplus L, \qquad c_1({V}) + c_1(L) =0.
\eeq
Here $V$ is a rank four spectral cover bundle descending from ${{\cal C}^{(4)}}$ via a spectral line bundle ${\cal N}_4$ with $c_1({\cal N}_4)  \in H^2({{\cal C}^{(4)}}; {\mathbb Z})$ as
\beq
  V = \pi_{4*} {\cal N}_4.
\eeq
$L$ is a line bundle that formally derives as the push-forward of ${\cal N}_1$ on ${\cal C}^{(1)}$. Note that this includes the special case where $c_1(L)=0=c_1(V)$, in which case the structure group of ${W}$ is just $SU(4)$.

To be precise, ${\cal N}_4$ is given by \eqref{linebundle} for $n=4$ and with $\eta$ replaced by $\tilde \eta =  \eta - c_1(S)$, and $c_1(V) \equiv \zeta = - c_1(L)  \in H^2(S; \mathbb Z)$, i.e.
\beq \label{linebundlesu4}
  \bal
    c_1(\mathcal{N}^{(4)})&= \frac{r^{(4)}}{2} + \gamma^{(4)}_u + \frac{1}{4} \pi^*_4 \zeta \\
    &= \left({\textstyle  1+4\lambda }\right)\,\sigma +   
       \left({\textstyle \frac{1}{2} - \lambda}\right) \pi_4^\ast \tilde\eta + \left( {\textstyle - \frac{1}{2}+ 4
       \lambda} \right) \pi_4^\ast c_1(S) + {\textstyle \frac{1}{4}}\, \pi_4^\ast \zeta.
  \eal
\eeq
For ${\cal N}_1$ one can formally set $n=1$, $\lambda  = \frac12$ so that $\frac{r^{(1)}}{2} + \gamma^{(1)}_u = 0$ and
\beq
  c_1({\cal N}_1) =  - \pi^*_1 \zeta. 
\eeq

Integrality of $c_1({\cal N}_4)$ requires that 
\beq \label{Nintegerb}
    4 \,\lambda  \in {\mathbb Z}\ , \qquad 
    \left({\textstyle \frac{1}{2}-\lambda}\right)\,\tilde \eta - {\textstyle \frac{1}{2}} c_1(S) + {\textstyle \frac{1}{4}}\, \zeta   \in  H^2(S; {\mathbb Z})\ .
\eeq
In particular, and in contrast to some claims in the recent F-theory literature, if the first Chern class $c_1(S)$ is odd it is not possible to set $\lambda =0$ and refrain from switching on universal gauge flux $\gamma$ altogether. This is the analogue of the Freed-Witten quantization condition in D-brane models \cite{Freed:1999vc}, which forces one in certain cases to include non-trivial gauge flux for quantum consistency of the construction.


\subsubsection*{Matter}
The group theoretic embedding of this $S[U(4) \times U(1)_X]$ bundle into $E_8$ is very similar to the procedure outlined in \secref{hyper}. The Cartan generators of ${W}$ are identified as $(1,1,1,1,-4) \subset SU(5)_{\perp}$ so that $V$ and $L$ are assigned $U(1)_X$ charges $+1$ and $-4$ respectively. The breaking pattern
\beq \label{breakingSU5}
  \bal
    E_8 {\,\,} & \longrightarrow {\,\,\,} SU(4) \times SU(5) \times U(1)_{X} \\
    {\bf 248} & \mapsto
    \left\{\begin{array}{l}
      ({\bf 15},{\bf 1})_0 \\
      ({\bf 1}, {\bf 1})_0 + ({\bf 1},{\bf 10})_{-4} + ({\bf 1}, \ov{\bf 10})_{4} + ({\bf 1}, {\bf 24})_0 \\
      ({\bf 4},{\bf 1})_{5} + ({\bf 4}, \ov{\bf 5})_{-3} + ({\bf 4},{\bf 10})_{1} \\
      (\ov{\bf 4},{\bf 1})_{-5} + (\ov{\bf 4}, {\bf 5})_{3} + (\ov{\bf 4},\ov{\bf 10})_{-1} \\
      ({\bf 6},{\bf 5})_{-2} + ({\bf 6},\ov{\bf 5})_{2}
    \end{array}\right\}
  \eal
\eeq
then immediately implies that the $SU(5)_{\GUT}$ matter is associated with the bundle representations displayed in \tableref{spec_E6}. Recall that absence of the ${\bf 10}_{-4}$ has been ensured by construction, see the discussion after \eqref{10curveb}.

\begin{table}[ht]
  \renewcommand{\arraystretch}{1.3}
  \centering
  \begin{tabular}{c|c|c}
    $SU(5)\times U(1)_X $ & bundle & SM particles$_\big.$ \\
    \hline \hline
    $ ({\bf 10},{\bf 1})_{1}$ & $V$  & $(q_L,u^c_R,e^c_R)$$^\big.$ \\
    $( {\bf 10},{\bf 1})_{-4}$ & $L$  &$-$$_\big.$ \\
    \hline
    $(\overline {\bf 5},{\bf 1})_{-{3}}$ & $V \otimes L$ & $(d^c_R,l_L)$$^\big.$ \\
    $({\bf \ov 5},{\bf 1})_{2}$ & $\bigwedge^2 V$  & $[(H_u, H_d)+(\ov H_u, \ov H_d)]$$_\big.$ \\
    \hline
    $({\bf 1},{\bf 1})_{{5}}$ & $V \otimes L^{-1}$ &  $\nu^c_R$$^\big.$
  \end{tabular}
 \caption{\small Dictionary between the standard model representations and bundles for the breaking $E_8 \longrightarrow SU(4) \times \big[ SU(5) \times U(1)_{X} \big]$.}
 \tablelabel{spec_E6}
\end{table}

The remaining and relevant ${\bf 10}_1$ is now given by
\beq \label{coho_10b}
  {\rm Ext}^i(\iota_* {\cal O}_{\sigma};  { j_4}_* {\cal N}_4 ) = H^{(i-1)} ({\cal C}^{(4)} \cap \sigma; {\cal N}_4 \otimes K_S |_{{\cal C}^{(4)} \cap \sigma}), \quad i = 1,2,
\eeq
with chiral index
\beq \label{chi10-split}
  \bal
    \chi_{\bf 10} &= \chi(\ccP_{10}, {\cal N}_4 \otimes K_S |_{\ccP_{10}} ) \\
     &= [\ccP_{10}]\cdot \bigl(\gamma^{(4)}_u+{\textstyle \frac{1}{4}}\pi^*_4 \zeta \bigr)  \\
     &= \big( - \lambda \tilde \eta + {\textstyle \frac14} \zeta \big) \cdot_S \underbrace{(\tilde \eta - 4 c_1(S)}_{P_{10}}).
  \eal
\eeq
To compute the ${\bf \ov 5}_m$ one has to appreciate that the matter curve ${\cal P}_5$ given in \eqref{classPH} is exactly of the form \eqref{Pab} for $C_1={\cal C}_4$, $C_2={\cal C}_1$ and $C_R = 2 \pi^* c_1(S)$. In particular our formula  \eqref{coho_ab}  for the representation of $V \otimes L$ applies and the chiral multiplets are counted by
\beq \label{coho_5m}
  \bal
  {\rm Ext}^i&({j_1}_* {\cal N}_1^{\vee};  { j_4}_* {\cal N}_4 ) \qquad\qquad\qquad\qquad\qquad i = 1,2 \\
  & {} = H^{(i-1)} ({\ccP}_m; {\cal N}_4  \otimes  {\cal N}_1  \otimes K_S \otimes K_X^{-\frac12} \otimes {\cal O}(-R/2) |_  {{\ccP}_m}  ).
  \eal
\eeq
Here ${\cal N}_1 ^{\vee}$ on the right-hand side of \eqref{coho_5m} is defined by replacing $-\zeta$ by  $\zeta$. One can then compute the corresponding index as
\beq \label{chi5-split}
  \bal
    \chi_{{\bf \ov 5}_m} &= \chi ({\ccP}_m, {\cal N}_4  \otimes  {\cal N}_1  \otimes K_S |_  {{\ccP}_m}  ) \\
    &= [\ccP_m] \cdot  \bigl(\gamma_u + {\textstyle \frac{1}{4}}\pi^*_4 \zeta  - \pi_1^*\zeta\bigr) \\
    &= \lambda \left( -\tilde\eta^2 + 6\tilde\eta c_1(S) -8 c^2_1(S)\right) + \textstyle \frac{1}{4} \zeta\, ( -3\tilde\eta +6 c_1(S)).
  \eal
\eeq
The number of Higgses again necessitates a more in-depth treatment analogous to the one around eqn.~\eqref{coho_5}, but their index is simply
\beq \label{chi5-splithiggs}
  \bal
    \chi_{{\bf \ov 5}_H} &=  [\ccP_H] \cdot  \bigl(\gamma_u + {\textstyle \frac{1}{4}}\pi^*_4 \zeta  \bigr)  \\
    &= \lambda \left( -2\tilde\eta c_1(S) +8 c^2_1(S)\right) + \textstyle \frac{1}{4} \zeta\, ( 4\tilde\eta -10  c_1(S))
  \eal
\eeq
for the Higgs curve $[\ccP_H]$ displayed in \eqref{classPH}. In fact one can easily convince oneself that 
\beq
  \chi_{\bf 10} = \chi_{{\bf \ov 5}_m} + \chi_{{ \bf \ov 5}_H},
\eeq
as required for a consistent gauge bundle giving rise to an anomaly-free spectrum.

Interestingly, the spectrum in \tableref{spec_E6} contains the new modes ${\bf 1_5}$, which were not present before the split $SU(5)_{\perp} \rightarrow S[U(4) \times U(1)_X]$. They emerge from the decomposition of the adjoint of $SU(5)_{\perp}$ as in 
\beq \label{24_dec}
  {\bf 24} \mapsto {\bf 15_0} + {\bf 1_0} + {\bf 4_5} + {\bf \ov 4_{-5}} .
\eeq
Since these Standard Model singlets have the correct $U(1)_X$ charge to participate in the coupling ${\bf \ov5_m}\,  {\bf 5_H}\, {\bf 1}$ their physical interpretation is as candidates for right-handed neutrinos \cite{Tatar:2009jk}. 
It is clear from their appearance in \eqref{24_dec} that they arise in the ${\cal C}^{(4)} - {\cal C}^{(1)}$ sector, but unlike the ${\bf \ov 5}_m$ they are localized away from the GUT brane $S$. The associated matter curve should thus be precisely the curve ${\cal P}_{\nu}$ which we introduced in equ.~\eqref{Pnu}. We can rewrite ${\cal P}_{\nu}$ somewhat redundantly as in \eqref{Pab} with $C_1 = {\cal C}_1$,  $C_2 = {\cal C}_4$, $C_R = {\cal C}_4 - 2 \,\pi^* c_1(S)$. Following our general logic the relevant cohomology groups are then
\beq \label{coho_nu}
  \bal
  {\rm Ext}^i&({j_1}_* {\cal N}_1;  { j_4}_* {\cal N}_4 ) \qquad\qquad\qquad\qquad\qquad i = 1,2 \\
  & {} = H^{(i-1)} ({\ccP}_{\nu}; {\cal N}_4  \otimes  {\cal N}_1^{\vee}  \otimes K_S \otimes K_X^{-\frac12} \otimes {\cal O}(- R/2) |_  {{\ccP}_{\nu}}  )  .
  \eal
\eeq
The chiral multiplicity derives from the index
\beq \label{chi-nu}
  \bal
    \chi_{\nu_R^c} &= \int_{{\cal P}_{\nu}} \gamma_4 + \frac14 \pi_4^*\zeta + \pi_1^*\zeta \\
    &= 2 c_1(S) \cdot \left(4 \,  \lambda  \, c_1(S) - \lambda \, \tilde \eta + \frac54\,  \zeta \right).
  \eal
\eeq
Since these modes are localized away from the GUT divisor $S$, it is
not entirely clear, though, whether in the description of  such global modes there arise additional subtleties 
not captured by the local spectral cover construction.

\subsection{GUT symmetry breaking and 3-brane tadpole}\seclabel{split_GUTbreaking}
Our final task is to incorporate the breaking of $SU(5)_{\GUT}$ into the split spectral cover construction. Again we face the problem that the straightforward implementation of a line bundle $L_Y$ on $S$ makes it hard to eliminate the exotic states in the $({\bf 3},{\bf 2})$. Not dissimilar to the approach in \secref{hyper} the solution is a specific twist of the bundle $L_Y$ with the spectral cover bundle.

Let us define the bundles ${\cal V}$, ${\cal L}_Y$ and ${\cal L}$ by
\beq \label{twist}
  V   = {\cal V} \otimes  {\cal L}_Y^{-1/5}, \qquad
  L   = {\cal L} \otimes {\cal L}_Y^{4/5}, \qquad
  L_Y = {\cal L}_Y^{1/5}.
\eeq
From \eqref{splitting} one can read off the hypercharge $q_Y$ of the individual MSSM representations descending from the various GUT multiplets. This gives an extra factor of $L_Y^{q_Y}$ in the bundle assignments of \tableref{spec_E6}. With the help of \eqref{twist} one then finds that the MSSM matter is counted by the cohomology groups associated with the bundles in \tableref{spec_E6b}. Note the appearance of integer powers despite the fractions in \eqref{twist}. 

To summarize, the gauge flux 
\begin{itemize}
  \item on the matter branes is described by an $S[U(4) \times U(1)_X]$ bundle ${\cal V} \oplus {\cal L}$ with $c_1({\cal V}) = - c_1({\cal L}) = \zeta \in H^2(S; \mathbb Z)$, and the flux
  \item on the GUT divisor $S$ is given the line bundle ${\cal L}_Y \in H^2(S; \mathbb Z)$.
\end{itemize}

\begin{table}[htb]
  \renewcommand{\arraystretch}{1.3}
  \centering
  \begin{tabular}{c|c|c}
    $SU(3)\times SU(2) \times U(1)_X  \times U(1)_Y$ & bundle & SM part.$_\big.$ \\
    \hline
    \hline
    $ ({\bf 3},{\bf 2})_{1_X, 1_Y}$ & ${\cal V}$  & $Q_L$$^\big.$ \\
    $ ({\bf \ov 3},{\bf 1})_{1_X, -4_Y}$ & ${\cal V} \otimes {\cal L}_Y^{-1}$  & $u^c_R$ \\
    $ ({\bf 1},{\bf 1})_{1_X, 6_Y}$ & ${\cal V} \otimes {\cal L}_Y$  & $e^c_R$$_\big.$ \\
    \hline
    $(\overline {\bf 3},{\bf 1})_{-{3}_X, 2_Y}$ & ${\cal V} \otimes {\cal L} \otimes {\cal L}_Y  $ & $d^c_R$$^\big.$ \\
    $( {\bf 1},{\bf 2})_{-{3}_X, -3_Y}$ & ${\cal V} \otimes {\cal L}$ & $  l_L $$_\big.$ \\
    \hline
    $({\bf 3},{\bf 1})_{-2_X,-2_Y}$ & $\bigwedge^2 {\cal V}$  & $-$$^\big.$ \\
    $({\bf 1 },{\bf 2 })_{-2_X,3_Y}$ & $\bigwedge^2 {\cal V} \otimes {\cal L}_Y^{-1}$  & $ H$$_\big.$ \\
    \hline
    $({\bf 1},{\bf 1})_{{5}_X,0_Y}$ & ${\cal V} \otimes {\cal L}^{-1}  \otimes {\cal L}_Y^{-1}$ &  $\nu^c_R$$^\big.$
  \end{tabular}
  \caption{\small Dictionary between the standard model representations and bundles for the breaking in the twisted split spectral cover construction.}
  \tablelabel{spec_E6b}
\end{table}

Finally we give the complete expression for the 3-brane tadpole \eqref{D3a}. First we need to evaluate the traces \eqref{traces W_1} over the spectrum in  \tableref{spec_E6b}. This yields
\beq
\frac{1}{4 (2 \pi)^2} \,  {\rm tr} \ov F^2 =  {\rm ch}_2({\cal V}) + {\rm ch}_2({\cal L})  + 2 \,  {\rm ch}_2 ( {\cal L}_{Y} ) + c_1({\cal L} ) \, c_1({\cal L}_Y)  .
\eeq
Correspondingly the flux part in \eqref{D3a} is
\beq
  \bal
    - \frac12 G^2 = {} & {} \frac{1}{2}\big( \textstyle - \lambda^2 n \tilde \eta(\tilde \eta - 4 c_1(S)) + \left(\frac{1}{4} +1\right) \zeta^2 \\
    &{} + 2 c_1^2({\cal L}_Y) - 2 \zeta c_1({\cal L}_Y) \big). 
  \eal
\eeq
For the curvature dependent part the factor $\chi_{SU(5)}$ splits into the sum of two   $\chi_{SU(n)}$ pieces with $n=4, \eta \rightarrow \tilde \eta$ and $n=1, \eta\rightarrow c_1(S)$. The latter gives no contribution, and the final result is
\beq \label{N3-split-gen}
  \bal
    N_{3} ={} & \frac{\chi^*(Y)}{24} - \frac{615}{2} \int_S c_1^2(S) - 15 \int_S  (  \eta^2 - 9 \, \eta\,  c_1(S)) \\
    &{} + \left(\frac12 - 2\,  \lambda^2\right) \int_S \tilde \eta \,  (\tilde \eta - 4 \, c_1(S)) \\
    &{} + \frac{5}{8} \zeta^2  +  \,  c_1^2({\cal L}_Y) -  \, \zeta \, c_1({\cal L}_Y),
  \eal
\eeq
where $\chi^*(Y)$ is the Euler characteristic \eqref{smooth_chi} evaluated for the 
base $B$.
That this $N_3$ is indeed an integer in concrete examples is a non-trivial consistency check.
As a further cross-check we have also applied the approach developed in \cite{Andreas:1999ng, Andreas:2009uf} for the non-factorized spectral covers to the factorized case. This requires in particular a careful treatment of the intersection locus of the ${\cal P}_m$ and ${\cal P}_H$ curves. 
As for the unfactorized cover this gives an ansatz with some universal unfixed parameters, which in particular must be valid for F-theory models with a heterotic dual.
Comparison with the heterotic dual, if available, is sufficient to fix these parameters. 
Here we assumed that the heterotic dual is described by  the product of the above two spectral covers of rank 4 and 1, respectively.
Due to its technical character we refrain from presenting this analysis here, but we stress that its result confirms the above value computed by our methods.

\subsection{Supersymmetry condition and stable extensions}\seclabel{SUSY_sec}
In F/M-theory the four-form flux $G$ preserves ${\cal N}=1$ supersymmetry if it is an element of $H^{(2,2)}(Y)$ and primitive, i.e.
\beq \label{JwedgeG}
  J_Y \wedge G = 0.
\eeq
These two constraints translate as follows into conditions on the spectral line bundle ${\cal N}$ under the dictionary \eqref{FG} \cite{Donagi:2008ca, Donagi:2009ra}. Since the two-cycles $\omega^I$ are dual to $(1,1)$ forms, the gauge flux $F$ and thus $\gamma$ must be elements of $H^{(1,1)}(S)$. This is the familiar F-term supersymmetry condition on the gauge flux on 7-branes. The so-called universal \cite{Donagi:2009ra} flux  \eqref{gammau} is by construction a $(1,1)$ form. The same is of course true for our generalization including a non-trivial $\zeta \in Pic(S)$. In general there will be extra fluxes which satisfy this constraint only for special values of the complex structure moduli.

Further one can evaluate the primitivity condition \eqref{JwedgeG} as equivalent to \cite{Donagi:2008ca, Donagi:2009ra} 
\beq
  \iota^* J \wedge \pi_{n*}\gamma = 0
\eeq
on $S$, which is readily recognized as the usual D-term supersymmetry condition for zero matter VEVs. Here $\iota^*J$ is the pullback of the K\"ahler form on $Y$. Clearly for $SU(n)$ bundles this is automatically satisfied because $\pi_{n*} \gamma = 0$ or equivalently $c_1(V)=0$. For  $U(n)$ bundles, by contrast, a D-term constraint remains. In principle, also the hyperflux bundle ${\cal L}_Y$ is subject to a D-term condition. Since a massless $U(1)_Y$ requires this flux to be localized on cycles which are trivial on the ambient space the associated D-term vanishes automatically. 

The field theoretic interpretation is the well-known phenomenon that the appearance of a massive $U(1)$ induces a field-dependent Fayet-Ilopoulos term. Concretely in our construction of $S[U(4) \times U(1)]$ bundles the Fayet-Iliopoulos term for $U(1)_X$ is given by
\beq
  \mu({V}) = \int_S  {\iota}^*J \wedge \zeta  = - \mu({L}) ,
\eeq
where for simplicity we are ignoring the twist by ${\cal L}_Y$ as it plays no role for the D-term. The full D-term is of course
\beq
  \sum_i q_i |\Phi_i|^2 + \mu,
\eeq
where $\Phi_i$ denote the scalars of charge $q_i$ under $U(1)_X$. The D-flatness condition thus stabilizes one linear combination of the charged scalars and the K\"ahler moduli. The usual two strategies are possible: Either one ensures that $\mu({V})=0$ inside the K\"ahler cone. In this case all fields charged under $U(1)_X$ must have zero VEV or their VEVs must cancel appropriately. Alternatively one can cancel $\mu$ against the vacuum expectation value of some scalars. For phenomenological reasons none of the MSSM fields should acquire a VEV as this would break the MSSM gauge group. An exception are the fields ${\bf 1_{5}}$ counted by ${\rm Ext}^1(L;V)$. In \secref{sec_Bundle} these modes were identified as candidates for right-handed neutrinos \cite{Tatar:2009jk}. From a mathematical point of view they allow for an interpretation as the recombination moduli for the bundles ${V}$ and $ L$. Giving them a VEV corresponds to forming an extension
\beq
  0 \rightarrow {V} \rightarrow W \rightarrow L \rightarrow 0.
\eeq
This requires that $\mu({V}) < 0$ for the extension to have a chance to be stable. In this process the $S[U(4) \times U(1)]$ bundle ${ V} \oplus { L}$ actually recombines into an $SU(5)_{\perp}$ bundle. As was stressed already in \cite{Tatar:2009jk} this higgses the massive $U(1)_X$.

\section{An explicit F-theory \boldmath$SU(5)$ GUT example}\seclabel{sec_geometry}

In this section we explicitly describe the Calabi-Yau fourfold geometries on which we will compactify F-theory to construct an $SU(5)$ GUT model. The specification of the brane fluxes will allow us to realize a three-generation $SU(5)$ GUT model. In \secref{FanodP7trans} we study a $dP_7$ transition of the Fano threefold $\bbP^4[4]$, and introduce some toric tools to analyze the resulting threefold $B$ efficiently. We are then in a position to construct the complete-intersecting Calabi-Yau fourfold $Y$ which admits $B$ as a base in \secref{complete_inters_ex}. Toric geometry allows us to resolve an $SU(5)$ degeneration over the del Pezzo surface in $B$ as described in \secref{toric_resolution_ex}. The fluxes on the 7-branes will then be constructed using the the split spectral cover construction of \secref{split_covers} and the $S[U(4) \times U(1)_X]$ bundles of \secref{sec_Bundle}.

\subsection[{Fano threefold $\bbP^4[4]$ and its del Pezzo 7 transition}]{Fano threefold \boldmath$\bbP^4[4]$ and its del Pezzo 7 transition}\seclabel{FanodP7trans}
In this section we introduce the base $B$ for a concrete Calabi-Yau fourfold geometry on which we will realize an $SU(5)$ GUT model. More concretely, we will generate a $dP_7$ surface by performing a del Pezzo transition starting with the Fano threefold $\bbP^4[4]$. We thus follow the strategy outlined in \secref{CompactGeometries}. However, it will be crucial to also introduce the corresponding toric description of the transitioned base $B$. This will allow us in \secref{complete_inters_ex} to construct the elliptically fibered Calabi-Yau fourfold as a complete intersection in a six-dimensional toric ambient space.

Let us thus start with the simple Fano threefold hypersurface $\mathbb{P}^4[4]$, given by $f_4(y_1,\ldots,y_5)=0$, which is the three-dimensional analog of the two-dimensional $dP_6=\bbP^3[3]$. It admits $45$ independent complex structure deformations, as can be checked by counting independent coefficients in $f_4$. One also computes that the Euler characteristic of the base is $\chi = -56$. By fixing some of the complex structure deformations one can generate a del Pezzo $7$ singularity, which is then blown up into a $dP_7$ surface of finite size. To achieve this one tunes the hypersurface constraint $\bbP^4[4]$ such that it takes the form
\beq \label{dP7_sing}
  P_{\rm base} = f_4+ y_4 \, f_3+y_5 \, g_3 + y_4^2\, f_2 + y_5^2\, g_2 + y_4 y_5\, h_2 ,   
\eeq
where $f_n,g_n,h_n$ are generic polynomials of degree $n$ in the three coordinates $(y_1,y_2,y_3)$. One easily shows that the manifold described by \eqref{dP7_sing} is singular on the $\bbP^1$ given by $(0,0,0,y_4,y_5)\sim \lambda (0,0,0,y_4,y_5)$ for $\lambda \in \bbC^\times$. This $\bbP^1$ can be blown up into a  $dP_7$ surface of finite size by including a new coordinate $w$ as
\beq \label{Pbase_dP7}
  P_{\rm base} = w^2\, f_4+ w (y_4 \, f_3+  y_5 \, g_3 ) + y_4^2\, f_2 + y_5^2\, g_2 + y_4 y_5\, h_2 , 
\eeq
and an additional scaling relation $(y_1,y_2,y_3,y_4,y_5,w) \sim (y_1,y_2,y_3,\lambda y_4,\lambda y_5,\lambda w)$. The $dP_7$ is obtained by setting $w=0$ in \eqref{Pbase_dP7} and thus given by 
\beq \label{dP_short}
  y_4^2\, f_2 + y_5^2\, g_2 + y_4 y_5\, h_2 =0 .
\eeq
Now one distinguishes two patches for which either $y_4=1,g_2\neq 0$ or $y_5=1,f_2\neq 0$. For example, in the latter one can multiply \eqref{dP_short} by $f_2$ and introduce the coordinate $p=y_4 f_2$ such that \eqref{dP_short} becomes
\beq \label{non-gendP7}
  P_{dP_7} = p^2 + p\, h_2 + g_2 f_2 = 0  .
\eeq
Only the scaling $(y_1,y_2,y_3,y_4,y_5) \sim (\lambda y_1,\lambda y_2,\lambda y_3, y_4, y_5)$ remains since we have fixed one of the scalings by setting $y_5=1$. Consistently with the fact that \eqref{non-gendP7} is a $dP_7$ surface the coordinates $(y_1,y_2,y_3,p)$ have weight $(1,1,1,2)$ under this scaling. Moreover, this $dP_7$ is non-generic since the $p$-independent term takes the special form $Q_4 = g_2 f_2$ and is not a generic quartic polynomial. In fact, this non-generic $dP_7$ has only $3$ complex structure deformations as opposed to $6$ for a generic $dP_7$.

Let us stress that the transition is best described using toric geometry which directly yields the resolved base \eqref{Pbase_dP7}. It is not hard to represent the ambient space $\bbP^4$ by a toric polyhedron given by the points $\nu_1,\ldots,\nu_5$ in \tableref{Base1}. After the del Pezzo transition the new base $B$ will admit an additional point $\nu_6$ and the new polyhedron is then spanned by the complete set of points listed in \tableref{Base1}.

\begin{table}[ht]
  \centering
  \begin{tabular}{c|r@{\,$=$\,(\,}r@{,\;\;}r@{,\;\;}r@{,\;\;}r@{\,)\;\;}|c|c} 
    polyhedron &\multicolumn{5}{c|}{vertices} & coords  & divisors$_\big.$ \\
    \hline\hline
    & $N_0$ & $0$ & $0$ & $0$ & $0$ & $y_0$  & $-4H-2X$$^\big._\big.$ \\ \hline
    $\nabla(B)$ 
    & $\nu_1$  & $-1$ & $-1$ & $-1$ & $-1$ & $y_1$  & $H$$^\big.$ \\
    & $\nu_2$  & $1$ & $0$ & $0$ & $0$ & $y_2$  & $H$\\
    & $\nu_3$ & $0$ & $1$ & $0$ & $0$ & $y_3$  & $H$\\
    & $\nu_4$ & $0$ & $0$ & $1$ & $0$ & $y_4$  & $H+X$\\
    & $\nu_5$ & $0$ & $0$ & $0$ & $1$ & $y_5$  & $H+X$\\
    & $\nu_6$ & $0$ & $0$ & $-1$ & $-1$ & $w$  & $X$$_\big.$\\
    \hline
    \multicolumn{7}{c|}{}    & $c_1 = X+H$$^\big.$
  \end{tabular}
  \caption{\small Toric variety of $\bbP^4$ with divisor $s=0$ corresponding to blow-up $dP_7$.}
  \tablelabel{Base1}
\end{table}

Recall that the hypersurface conditions $\{y_i = 0\},\{w=0\}$ define toric divisors in the ambient space which we will denote by $D_{y_i},D_w$. There are two linear relations among the six points $\nu_i$ and hence among the homology classes of the six divisors $D_{y_i},D_w$. This allows us to reduce the set of divisor classes to two independent classes $H$ and $X$, where $H$ is the analog of the hyperplane class of $\bbP^4$ and $X$ is the class of the blow-up divisor.

Using standard methods of toric geometry, one can determine the intersection numbers and Chern classes of the base $B$. First one determines the star triangulations of $\nabla$ in which every vertex contains the origin.\footnote{This can be done using the program TOPCOM \cite{Rambau:TOPCOM-ICMS:2002}.} This determines the intersection numbers of the four-dimensional toric ambient space. Then one needs to restrict these numbers to the hypersurface $B$ in the class $4 H+2 X$. The toric divisors $D_{y_i},D_w$ restrict to divisors of the hypersurface such that one can determine the triple intersections of this threefold by restriction of the intersections of the toric ambient space. By abuse of notation we will denote the divisors restricted to the hypersurface also by $D_{y_i},D_w$ and $H,X$. Explicitly one finds that\footnote{The computation of the intersection numbers on a hypersurface or complete intersection is performed by the Maple package Schubert by S.~Katz and S.~A.~Str{\o}mme.}
\beq
\label{intformb}
  H^3 = 0 , \qquad 
  H^2 X =2 , \qquad 
  H X^2 = 0 , \qquad
  X^3 = -2  .
\eeq
Let us note that there are negative intersection numbers, which indicates that $H,X$ are not the generators of the K\"ahler cone in which all volumes are positive. In fact, the K\"ahler cone is spanned by $K_1 = H$ and $K_2 = X+H$, such that $K_2^3 = 4$, $K_1^2 K_2 = 2$ and $K_1 K_2^2=4$. Expanding the K\"ahler form as $J=v^1 [K_1]+v^2 [K_2]$, all physical volumes will be positive in the K\"ahler cone $v^i>0$.

It is straightforward to check that the divisor $X$ has $\int_X c_1^2 = 2$ and Euler characteristic $\chi(X) = 10$. This is consistent with the fact that $X$ is a $dP_7$ surface. As we have stressed this $dP_7$ is not generic, since it arose by blowing up singularities over $\bbP^1$. This can be also inferred from the fact that $\chi(B) = - 34$ such that in the del Pezzo transition we have
\beq
  \Delta \chi = 22 = 2\, C_{E_6} - \chi(\bbP^1) ,
\eeq
where $C_{E_6}=12$ is the dual Coxeter number of $E_6$. The first term would appear in generic $dP_6$ transitions, while for a generic $dP_7$ transition one would find $\Delta \chi = 2\, C_{E_7}=36$. One 
thus arrives at a non-generic $dP_7$ surface, for which only an $E_6$ lattice is trivial in the base $B$. Torically realized transitions to  non-generically embedded del Pezzos have already appeared for Calabi-Yau threefolds in refs.~\cite{Grimm:2008ed, Blumenhagen:2008zz}.

\subsection{Constructing the complete-intersecting CY fourfold}\seclabel{complete_inters_ex}
In the following we specify the Calabi-Yau fourfold obtained as a complete intersection of two hypersurfaces in a six-dimensional toric ambient space as in \secref{complete_intersect_gen}. We then degenerate the elliptic fibration over the $dP_7$ surface and resolve the singularity torically. 

The construction of complete intersections in a toric ambient space is more involved than in the case of hypersurfaces \cite{Borisov-1993, batyrev-1994-1, batyrev-1994-2, Kreuzer:2001fu, Klemm:2004km}. This can be traced back to the fact that the intersection of the two hypersurfaces yielding the Calabi-Yau fourfold $Y$ has to be transversal in order that $Y$ is non-singular. This fact can be encoded in the toric polyhedron $\nabla$ describing the six-dimensional ambient variety, if there exists a split of $\nabla$ into two sets $\nabla_1$ and $\nabla_2$ such that the Minkowski sum $\nabla_1 + \nabla_2$ is a reflexive polyhedron.\footnote{Here $\nabla_1 + \nabla_2$ is the set consisting of the sums of each point in $\nabla_1$ with each point in $\nabla_2$. The polyhedron $\nabla_1+\nabla_2$ is reflexive if the origin is the only interior point of the polyhedron.} In case this condition is satisfied one calls the split $(\nabla_1,\nabla_2)$ a nef partition of $\nabla$ \cite{Borisov-1993, Kreuzer:2001fu}. In the following we will describe the construction of $(\nabla_1,\nabla_2)$ and the corresponding Calabi-Yau fourfold by focusing on concrete examples. 

A simple complete intersection can be constructed by considering an elliptic $\bbP_{123}[6]$ fibration over the Fano hypersurface $\bbP^4[4]$. Before discussing the construction of $Y$ in more detail, let us 
summarize the polyhedron in \tableref{P4[4]complete}. The points in $\nabla$ can be identified as follows. Firstly, recall that the polyhedron for $\bbP_{123}$ is spanned by the three toric points $\tilde \nabla=((-1,0),(0,-1),(3,2))$. These points appear in the first two rows of $\nabla$, and correspond to the coordinates $(y,x,z)$ of the elliptic fiber. As we will see explicitly below, the complete intersection is elliptically fibered since $\nabla$ contains the points $\nu_1,\nu_2,\nu_3$ which have the points $\tilde \nabla$ in the first two entries but are zero otherwise. This is in accord with the general arguments of ref.~\cite{Avram:1996pj}. The polyhedron of the base $\bbP^4$ appears in the last four rows of $\nabla$. It is not hard to check that the split of $\nabla$ into $(\nabla_1,\nabla_2)$
determines a valid nef partition.\footnote{This can be done by using the program PALP \cite{Kreuzer:2002uu}.}

\begin{table}[ht]
  \centering
  \begin{tabular}{c|r@{\,$=$\,(\,}r@{,\;\;}r@{,\;\;}r@{,\;\;}r@{,\;\;}r@{,\;\;}r@{\,)\;\;}|c} 
    nef-part. &\multicolumn{7}{c|}{vertices} & coords$_\big.$   \\
    \hline\hline
    & $N_0$ & $0$ & $0$ & $0$ & $0$ & $0$ & $0$ & $x_0$$^\big._\big.$ \\ \hline
    $\nabla_1$ & $\nu_1$ & $-1$ & $0$ & $0$ & $0$ & $0$ & $0$ & $y$$^\big.$  \\
    & $\nu_2$ & $0$ & $-1$ & $0$ & $0$ & $0$ & $0$ & $x$  \\
    & $\nu_3$ & $3$ & $2$ & $0$ & $0$ & $0$ & $0$ & $z$  \\
    & $\nu_4$ & $3$ & $2$ & $-1$ & $-1$ & $-1$ & $-1$ & $y_1$$_\big.$  \\ \hline
    $\nabla_2$ & $\tilde \nu_1$ &   $0$  &   $0$  &   $1$  &   $0$  & $0$ & $0$ & $y_2$$^\big.$ \\
    & $\nu_5$ &   $0$  &   $0$  &   $0$  &   $1$  &   $0$  &   $0$ & $y_3$ \\
    & $\nu_6$ &   $0$  &   $0$  &   $0$  &   $0$  &   $1$  &   $0$ & $y_4$ \\
    & $\nu_7$ &   $0$  &   $0$  &   $0$  &   $0$  &   $0$ &    $1$ & $y_5$$_\big.$ \\ \hline
  \end{tabular}
  \caption{\small The polyhedron for the complete-intersecting Calabi-Yau fourfold which is an elliptic fibration over $\bbP^4[4]$.}
  \tablelabel{P4[4]complete}
\end{table}

Let us now analyze the Calabi-Yau fourfold $Y$ defined by the data $(\nabla_1,\nabla_2)$ in more detail. It is given by a generic element of the class 
\beq
  \sum_{\nu_i \in \nabla_1} D_{x_i} \quad \cap \quad \sum_{\nu_j \in \nabla_2} D_{x_j} \ .
\eeq
where $x_i=(x,y,z,y_j)$ are the coordinates introduced in \tableref{P4[4]complete}. The two divisors correspond to the two hypersurface constraints defining $Y$. They can be given explicitly, in terms of the toric data if one determines the dual polyhedra to $(\nabla_1,\nabla_2)$. These dual polyhedra are the Newton polyhedra $(\Delta_1,\Delta_2)$ which obey~\cite{Cox:2000vi}
\beq \label{NablaDelta_dual}
  \langle \nabla_n ,\Delta_m \rangle \geq - \delta_{mn} .
\eeq 
Note that while $(\nabla_1,\nabla_2)$ only contain few points, the duals $(\Delta_1,\Delta_2)$ contain over one thousand points. This corresponds to the fact that the points of $(\nabla_1,\nabla_2)$ yield coordinates and toric divisors for the complete intersection, while the points in $(\Delta_1,\Delta_2)$ correspond to the monomials of the two hypersurface constraints. More precisely, the complete intersection is given by the two constraints 
\beq \label{complete_int}
  f_m = \sum_{w_k \in \Delta_m} c^{(m)}_{k} \prod_{n=1}^2 \prod_{\nu_i \in \nabla_n} x_i^{\langle \nu_i,w_k \rangle+\delta_{mn}}= 0 , \quad m=1,2 ,
\eeq
where $x_i=(x,y,z,y_j)$ are the coordinates introduced in \tableref{P4[4]complete}. The coefficients $\smash{c_k^{(m)}}$ parametrize the complex structure deformations of $Y$. 

One checks explicitly that $f_1=0$ is precisely the Tate form \eqref{Tate1} since $\nabla_1$ contains the points $(\nu_1,\nu_2,\nu_3)$ corresponding to the coordinates $(x,y,z)$ of the elliptic fiber in \eqref{Tate1}. The coefficients $a_n$ in the Tate form can be explicitly given in terms of the toric data. Let us first introduce the sets 
\beq \label{def-Ar}
  A_r = \{w_k \in \Delta_1:\, \langle \nu_3,w_k \rangle = r+1\} .
\eeq 
These are the elements in the Newton polyhedron which generate the monomials in the Tate form $f_1=0$ containing the power $z^r$. Hence, we can write 
\beq \label{toric_ai}
  a_r = \sum_{w_k \in A_r} c^{(1)}_{k} \prod_{n=1}^2 \prod_{\ \nu_i \in \nabla_n,\,i>3} \!\!\!\! y_i^{\langle \nu_i,w_k \rangle+\delta_{1n}} ,
\eeq
where we recall that $\nu_3$ corresponds to the $z$-coordinate of the elliptic fiber in \eqref{Tate1} and hence $a_r$ appears in front of $z^r$. Moreover, $f_2=0$ is the constraint for the base $\bbP^4[4]$. 

It is important to stress that our presentation is rather general and can be applied to other complete intersections accordingly. In particular, one can replace the base $\bbP^4[4]$ with the threefold obtained by blowing up a del Pezzo $7$ in $\bbP^4[4]$. In order to do that we have to use the polyhedron of \tableref{Base1} to obtain a Calabi-Yau fourfold $Y$ with nef partition as specified in \tableref{P4[4]completedP7}. Note that only a new divisor $(3,2,0,0,-1,-1)$ has been added which corresponds to the $dP_7$ when restricted to the base $B$ of $Y$. For this example one straightforwardly evaluates \eqref{toric_ai} and finds 
\beq
  a_r = \sum_{n=0}^r\, w^{r-n}\, \sum_{a=0}^n y_4^a\, y_5^{n-a}\, f^{(a,r)}_{r-n}(y_1,y_2,y_3)\ ,
\eeq
where $f^{(a,r)}_{r-n}$ are generic polynomials of degree $r-n$ in the coordinates $(y_1,y_2,y_3)$.

\begin{table}[ht]
  \centering
  \begin{tabular}{c|r@{\,$=$\,(\,}r@{,\;\;}r@{,\;\;}r@{,\;\;}r@{,\;\;}r@{,\;\;}r@{\,)\;\;}|c} 
    nef-part. &\multicolumn{7}{c|}{vertices} & coords$_\big.$   \\
    \hline\hline
    & $N_0$ & $0$ & $0$ & $0$ & $0$ & $0$ & $0$ & $x_0$$^\big._\big.$ \\ \hline
    $\nabla_1$ & $\nu_1$ & $-1$ & $0$ & $0$ & $0$ & $0$ & $0$ & $y$$^\big.$  \\
    & $\nu_2$ & $0$ & $-1$ & $0$ & $0$ & $0$ & $0$ & $x$  \\
    & $\nu_3$ & $3$ & $2$ & $0$ & $0$ & $0$ & $0$ & $z$  \\
    & $\nu_4$ & $3$ & $2$ & $-1$ & $-1$ & $-1$ & $-1$ & $y_1$  \\
    & $\nu_5$ & $3$ & $2$ & $0$ & $0$ & $-1$ & $-1$ & $w$$_\big.$ \\ \hline
    $\nabla_2$ & $\nu_1$ &   $0$  &   $0$  &   $1$  &   $0$  & $0$ & $0$ & $y_2$$^\big.$ \\
    & $\nu_5$ &   $0$  &   $0$  &   $0$  &   $1$  &   $0$  &   $0$ & $y_3$ \\
    & $\nu_6$ &   $0$  &   $0$  &   $0$  &   $0$  &   $1$  &   $0$ & $y_4$ \\
    & $\nu_7$ &   $0$  &   $0$  &   $0$  &   $0$  &   $0$ &    $1$ & $y_5$$_\big.$ \\ \hline
  \end{tabular}
  \caption{\small The polyhedron with $dP_7$.}
  \tablelabel{P4[4]completedP7}
\end{table}

\subsection{Resolving singularities in elliptic Calabi-Yau fourfolds}\seclabel{toric_resolution_ex}
Having constructed the Calabi-Yau fourfold as an elliptic fibration over the base $B$ with a del Pezzo surface, we next want to use toric geometry to degenerate the elliptic fiber of this surface to an $SU(5)$ singularity. Our strategy is to use the explicit expressions for the $a_i$ given in \eqref{toric_ai} and drop all monomials that would violate the $SU(5)$ form \eqref{TateSU5b}. Recall that in \eqref{TateSU5b} we have demanded that the $(a_1,a_2,a_3,a_4,a_6)$ contain the overall factors $(1,w,w^2,w^3,w^5)$ with coefficient functions $\mathfrak{b}_r(w,y_i)$. Hence, one has to drop the monomials in $a_r$ which admit powers of $w^k$ with $k<(0,1,2,3,5)$, respectively. 
Torically this is achieved by dropping points in the Newton polyhedron $\Delta_1$. One thus drops points of the sets $A_r$, defined in \eqref{def-Ar}, encoding the monomials in $a_r$, and obtains new sets  
\beq
  A_r^{SU(5)} \subset A_r  , \qquad \Delta_1^{SU(5)} \subset \Delta_1 . 
\eeq
Using the new Newton polyhedron in the Tate form $f_1 = 0$ in \eqref{complete_int} ensures that $Y$ will admit the singularity of the desired $SU(5)$ type. Clearly, one can use this method to generate also higher gauge groups on the divisor $S$ by imposing stronger constraints on the allowed monomials. Maximally, we can degenerate the elliptic fiber to $E_8$ by reducing to $\Delta_1^{E_8}$ dropping all monomials in the $a_r$ with powers lower than $(w,w^2,w^3,w^4,w^5)$.

Toric geometry can now be used to automatically resolve the singularities of the elliptic fibration. In order to do that one has to add new blow-up divisors, or, equivalently, new points to the polyhedron $\nabla=(\nabla_1,\nabla_2)$. More precisely, this can be done by determining the new duals of $(\Delta^{SU(5)}_1,\Delta_2^{SU(5)} = \Delta_2)$ via
\beq \label{NablaDelta_dualSU(5)}
  \langle \nabla^{SU(5)}_n ,\Delta^{SU(5)}_m \rangle \geq - \delta_{mn} 
\eeq 
just as in \eqref{NablaDelta_dual}. Since $\Delta^{SU(5)}\subset \Delta$ the polyhedron $\nabla^{SU(5)}=(\nabla^{SU(5)}_1,\nabla^{SU(5)}_2)$ will contain more points than the original polyhedron $\nabla$. The extra points $\tilde \nu_i$ are in $\smash{\nabla_1^{SU(5)}}$ and explicitly given by
\beq \label{blowup}
  \begin{array}{rr@{\,=\,(\,}r@{,\;\;}r@{,\;\;}r@{,\;\;}r@{,\;\;}r@{,\;\;}r@{\,)\;\;}}
    \nabla_1^{SU(5)}: & \nu_5       & 3 & 2 & 0 & 0 & -1 & -1, \\
                      & \tilde\nu_1 & 2 & 1 & 0 & 0 & -1 & -1, \\
                      & \tilde\nu_2 & 1 & 1 & 0 & 0 & -1 & -1, \\
                      & \tilde\nu_3 & 1 & 0 & 0 & 0 & -1 & -1, \\
                      & \tilde\nu_4 & \;0 & \;0 & \;0 & \;0 & -1 & -1,
  \end{array}
\eeq
where we have also recalled the divisor $\nu_5$ corresponding to the $dP_7$ in the base. The new points \eqref{blowup} together with the points in \tableref{P4[4]completedP7} define a new complete-intersecting fourfold $Y^{SU(5)}$, which admits a resolved $SU(5)$ fiber. Using these data one can again determine all relevant topological data for the fourfold such as intersection numbers and Chern classes. In particular, one finds that 
\beq 
  \chi(Y^{SU(5)}) = 918 .
\eeq 
This is in accord with the result obtained using the formulas for the 3-brane tadpole presented in \appref{Tadpoles_sec}.
One also readily derives the geometric data for the $SU(5)$ GUT model. One checks that the 
classes of the $\mathfrak{b}_n$ are $[\mathfrak{b}_n] =(6-n) H + X$.
The classes for the 
$\bf{10}$-curve \eqref{curve10} and $\bf{5}$-curve \eqref{curve5a} on the base $B$ are then given by
\beq
  [P_{10}] = (H+X)|_X \ , \qquad [P_5] = (8H+3X)|_{X} \ ,
\eeq
where the latter is to be taken with care since $P_5$ is singular. 
This geometry also realizes the ${\bf 10 \, 10 \, 5_H}$, ${\bf 10 \, \ov 5_m \, \ov 5_H}$ and ${\bf  5_H \, \ov 5_m \, 1}$ Yukawa couplings. These are respectively localized at the   $X \cdot (H+X) \cdot (2 H+X) = 2$ points of $E_6$ enhancement,  $X \cdot (H+X) \cdot (3H +X) = 4$ points of $D_6$ enhancement, and a non-vanishing number of points 
of $A_6$ enhancement in $B$.

Let us stress that one can use this technique to generate and resolve singularities also for other gauge groups realized on $S$ or some other toric divisor. In particular, one finds for the maximal $E_8$ case that one has to add the points $(3,2,n \vec{\mu}),\, n=1,...,6$, $(2,1,n \vec{\mu}),\, n=1,...,4$, $(1,1,n \vec{\mu}), n=1,2,3$, $(1,0,n \vec{\mu}), n=1,2$ and $(0,0,\vec{\mu})$, where $\vec{\mu}=(0,0,-1,-1)$. Finally, let us note that the general strategy to generate and resolve singularities in the complete-intersecting Calabi-Yau fourfolds is in accord with the results of refs.~\cite{Candelas:1996su, Candelas:1997eh} obtained for F-theory on elliptic Calabi-Yau threefold hypersurfaces.

\subsection{Gauge fluxes for a three-generation GUT model}\seclabel{sec_example}

\subsubsection*{The split spectral cover}
After this preparation we finally come to the explicit construction of our GUT model. Our first aim is to construct the Weierstrass model corresponding to the  $S[U(4)\times U(1)_X]$ spectral cover over the $SU(5)$ GUT brane $S$. The latter is chosen of course along the $dP_7$ surface $w=0$ specified by $c_1(N_{S/B}) = X$ in $B$, see \tableref{Base1}. We must first check whether it is indeed possible to find sections $c_i, d_i$ as in \eqref{split_sections} which satisfy the non-trivial relation \eqref{cconstr} between $c_0, c_1$ and $d_0$. Recall  that these are sections of line bundles corresponding to the classes
\beq
  \bal
    & [d_0] = c_1(S)  = -c_1(K_B) - c_1(N_S) = H, \\
    & [c_1] = \eta - 2 c_1(S) = 4H +X, \\
    & [c_0] = \eta - c_1(S) = 5H +X.
  \eal
\eeq

In the coordinates of \tableref{Base1} the most general ansatz for such sections is
\bea
  \bal
    d_0 &= p_1(y_1,y_2,y_3), \\
    c_1 &= w \,  p_4(y_1,y_2,y_3) + q_1(y_4,y_5) \, p_3(y_1,y_2,y_3), \\
    c_0 &=  w \, p_5(y_1,y_2,y_3) + r_1(y_4,y_5) \, q_4(y_1,y_2,y_3),
  \eal
\eea
where the $p_i, q_j, r_l$ denote arbitrary polynomials of the respective degree. This shows that the non-trivial factorization condition \eqref{cconstr} for the existence of a split spectral cover can indeed be met by suitable restrictions on these polynomials.

The classes of the  matter curves  are
\beq \label{matter_classes1}
  \bal
    {}[P_{10}] &= -c_1(K_{B})|_X = (H+X) |_X , \\
    {}[P_{H} ]&= -5c_1(K_{B})|_X - 3 c_1(N_{S/B})  = (5 H + 2 X)|_X,    \\
    {}[P_m ]&=  -3c_1(K_{B})|_X - 2 c_1(N_{S/B})  = (3 H +  X)|_X.
  \eal
\eeq
As mentioned already in \secref{toric_resolution_ex},
this geometry indeed realizes the ${\bf 10 \, 10 \, 5_H}$ and ${\bf 10 \, \ov 5_m \, \ov 5_H}$ Yukawa couplings 
at $2$ points of $E_6$ enhancement and  $4$ points of $D_6$ enhancement, respectively.

To gain  a better understanding of the matter curves on $S$ let us express the pullbacks in \eqref{matter_classes1} as elements of $H^2(S; \mathbb Z)$. As discussed in \secref{FanodP7trans} only an $E_6$ sublattice of the $dP_7$ is trivial on $B$. From $H|_X = c_1(S)$ it is clear that
\beq
  C_1 = H|_X =  - f = 3 l - \sum_i E_i,
\eeq 
which constitutes the canonical genus 1 curve on the del Pezzo $S$. On the other hand, also $C_2 = (H+X)|_X$ is a non-trivial curve on $S$ with  
\beq
  \chi(C_2) =  2, \qquad C_1 \cdot C_2 = 2.
\eeq
It is now easy to convince oneself that up to isomorphisms the only degree-2 curves of genus 0 are given by $l-E_i$. At first sight, the general ansatz $C= a_l l + \sum_i a_i E_i$ also allows for $C= 2l - E_i - E_j - E_k - E_l$. However, it is always possible to bring this in the form $l-E_i$ via the involution $\smash{I_{{\cal B}_7}^{(6)}}$ acting on $dP_7$ as classified in Table~22 of \cite{Blumenhagen:2008zz}. Let us therefore take for definiteness
\beq
  C_2 =  (H+X)|_X = l-E_7.
\eeq
Cohomologically this means
\beq
  [P_{10}]= l-E_7 ,\qquad [P_H] = -3 f + 2 (l-E_7), \qquad [P_m] = - 2 f + l - E_7.
\eeq

\subsubsection*{The bundle}
Let us first analyze the $S[U(4) \times U(1)_X]$ bundle on the matter branes, for the $SU(5)$ GUT model prior to gauge symmetry breaking. Specifically we need an ansatz for the spectral line bundle ${\cal N}_4$ given in \eqref{linebundlesu4}. It is given by
\beq
  \zeta =  (a X + b H) |_X, \qquad \lambda = \frac{x}{4}, \quad x \in \mathbb Z, \qquad \tilde \eta = 5 H +X.
\eeq
The integrality condition \eqref{Nintegerb}  can be explicitly evaluated as the constraint that
\beq
  a + 2 - x \in 4 {\mathbb Z}, \qquad b - 5x \in 4 {\mathbb Z}.
\eeq

The minimal phenomenological requirements are to have three chiral families of ${\bf 10}$ and ${\bf \ov 5}_m$ and no chiral excess of Higgs multiplets. This is equivalent to solving for $\chi_{\bf 10} = \pm 3$ and $\chi_{{\bf 5_H}} = 0$, where the overall sign of the chirality is merely a matter of convention. We can evaluate the expressions \eqref{chi10-split} and \eqref{chi5-split} with the above parametrization of the bundles in the present geometry as
\beq
  \chi_{\bf 10} = \frac{b-a-4x}{2}  = \pm 3, \qquad
  \chi_{\bf 5_H}=   -2 a + 5 b -  x =  0.
\eeq
A simple solution with the lower sign that is in agreement with the quantization condition is
\beq
  \zeta =( 10 X + 4 H) |_X, \qquad \lambda =0.
\eeq
This leads to 
\beq
  \chi_{\nu_R^c} = 5
\eeq
chiral generations as candidates for right-handed neutrinos.

On $S$ we furthermore choose the hypercharge flux corresponding to the $E_6$ root
\beq
  c_1({\cal L}_Y) = E_1 - E_2.
\eeq
Being trivial on the ambient space, this flux indeed leads to a massless $U(1)_Y$ upon GUT breaking. Second, it  restricts trivially to the matter curves $P_{10}$, $P_m$ and $P_H$ and thus does not affect the chiral index of our three-generation model. By contrast, as stressed several times, the computation of the non-chiral matter is more involved. In particular for the Higgs fields one would like to choose Wilson lines in such a way as to lead to one massless pair of MSSM Higgs doublets while making the triplets massive. For a  detailed discussion  on how to achieve this we refer to \cite{Blumenhagen:2008zz} but performing this analysis explicitly is beyond the scope of this paper.

\subsubsection*{D-term}
From the discussion after \eqref{intformb} we recall that the K\"ahler cone of $B$ is given by
\beq
  K_1 = H+X, \qquad K_2 = H.
\eeq 
After expanding $J = r_1 K_1 + r_2 K_2$ one finds that the Fayet-Iliopoulos term
\beq
  \mu({\cal V}) = \int_X J \wedge \zeta = -12 r_1 + 8 r_2
\eeq
can be set to zero within the K\"ahler cone. This corresponds to a solution to the D-flatness conditions on the locus of vanishing scalar fields. Alternatively it can be arranged that $ \mu({\cal V}) <0$ inside the K\"ahler cone, in which case a non-zero VEV for right-handed neutrinos alias extension moduli breaks $U(1)_X$ and forms an extension bundle, as summarized in \secref{SUSY_sec}.

\subsubsection*{3-brane tadpole}  
The curvature dependent part of the 3-brane tadpole can be evaluated for the split spectral cover by our simple method outlined in \secref{GeomTad} and \appref{Tadpoles_sec}.  Evaluating the first line and the $\lambda$-independent part of the second line of the general expression \eqref{N3-split-gen} for the manifold under consideration yields
\beq
  \frac{\chi(Y) }{24} = \frac{744}{24} = 31.
\eeq
The flux dependent 3-brane tadpole can be computed with the help of 
\beq
  \frac12 \pi_{4*} (\gamma_4^2) = \frac14(-a^2 + b^2) - 16 \lambda^2, \qquad \frac12 \zeta^2 = - a^2 + b^2.  
\eeq

For the above values and together with $c_1^2({\cal L}_Y) = -2$, $c_1({\cal L}_Y) \,  \zeta = 0$   the final result is
\beq \label{N3-ex}
  N_{3} = \frac{\chi(Y_4)}{24} - 105 - 2 = -76.
\eeq
Unfortunately, the bundle leads to considerable overshooting as a consequence of the comparatively small values for $ {\chi(Y)}/{24}$, i.e.~cancellation of the 3-brane tadpole requires the inclusion of anti-3-branes. We believe that this is an artifact  of the simple geometry we are considering here, and have evidence that this problem can be ameliorated by simple extensions of this setup \cite{progress}.

On the other hand, the fact that the right-hand side of \eqref{N3-ex}  is indeed an integer serves as an important and a non-trivial consistency check, in particular of the quantization of the bundles.

\subsection{Phenomenological aspects and outlook}
\label{sec_Pheno}
We have thus succeeded in constructing an explicit F-theory compactification with $SU(5)$ GUT symmetry broken to $SU(3) \times SU(2) \times U(1)_Y$ by hypercharge flux and three chiral generations of ${\bf 10}$ and ${\bf \ov 5_m}$. The $U(1)_Y$ has been arranged in such a way as to not intersect the ${\bf 10}$ and ${\bf \ov 5_m}$ matter curves so that the exact multiplicities of the MSSM matter descending from them are unaffected by GUT symmetry breaking. In this work we do not compute the exact matter spectrum, though, i.e. the vector-like states that are invisible to the chiral index. In particular such an analysis is required to determine the details of the Higgs sector and  is left to a future publication. 

The model incorporates the ${\bf 10 \, 10 \, 5_H}$ and the  ${\bf 10 \, \ov 5_m \, \ov 5_H}$ couplings, while potential dimension 4-proton decay operators ${\bf 10  \, \ov 5_m \, \ov 5_m}$ or ${\bf 10  \, \ov 5_H \, \ov 5_H}$  are absent by virtue of the extra massive $U(1)_X$ symmetry. As pointed out in \cite{Marsano:2009gv} the detailed form of the spectral cover used here leads to a problematic Higgs sector: The $H_u$ and $H_d$ both localize as a vector-like pair on a single curve $P_H$. Avoiding a high-scale $\mu$ term will therefore involve considerable fine-tuning. Also dimension-5 proton decay operators cannot be suppressed by a missing partner mechanism. As noted in \cite{Marsano:2009gv} both problems require a refinement of the factorization of the spectral cover such that $P_H$ splits into two distinct curves, which is an open challenge as of this publication.

Another important aspect of GUT models is gauge coupling unification.
In F-theory and Type IIB orientifold models with the breaking
of the GUT symmetry via a non-vanishing $U(1)_Y$ flux gauge coupling
unification only holds at leading order, as there exist $\alpha'$ corrections
due to the $U(1)_Y$ flux from the Chern-Simons term
\bea
\label{chernsim}
          S_{CS}=\mu_7 \,  \int_{\mathbb R^{1,3}\times S} C_0\wedge {\rm tr} (F^4)
       \; .
\eea
These have been computed in the framework
of $U(5)$ Type IIB orientifolds GUTs  
in  \cite{Blumenhagen:2008aw}, where also 
the presence of the line bundle ${\cal L}_a$ in the diagonal
$U(1)_a\subset U(5)$ was taken into account.
By a certain twisting, it was possible that all three
MSSM gauge couplings still unify after the corrections
\eqref{chernsim} have been taken into account.
Generically however the corrected gauge couplings do not unify exactly
but satisfy the relation
\bea
\label{relc}
        \frac{1}{\alpha_Y(M_s)}=\frac{1}{\alpha_w(M_s)} + \frac{2}{3\,
          \alpha_s(M_s)}\; ,
\eea
which could be reconciled with the running of the gauge couplings
if there exist a threshold $M_{3\overline 3}<M_X$ of the Higgs 
colour triplets $({\bf 3},{\bf 1})_{-\frac{2}{3}} +
({\bf\overline 3},{\bf 1})_{\frac{2}{3}}$ below the GUT scale.

Let us now address the situation for the F-theory GUT models
we were discussing in the previous sections. This is precisely
the computation carried out in  
\cite{Donagi:2008kj}. 
In proper F-theory models the interpretation of couplings analogous to (\ref{chernsim}) is less immediate due to the special role of the varying axio-dilaton. Our strategy is therefore to start from
(\ref{chernsim}), which is correct in the orientifold limit, and to extrapolate to F-theory by replacing the varying $g_s$ with the F-theory mass scale is $M_*^4=g_s^{-1}\ell_s^{-4}$. By duality with M-theory, the  latter is constant over the base $B$ (see e.g. \cite{Tatar:2009jk} for a recent discussion).

For the models under consideration,
the starting point is not $SO(32)$, but $E_8$, and
on the GUT divisor $S$ there is globally just the
non-vanishing $U(1)_Y$ flux $\overline{f}_Y$ of the 
line bundle ${\cal L}_Y$. The $U(1)_X$ flux is described
by a line bundle on the spectral covers ${\cal C}^{(4)}$ and ${\cal C}^{(1)}$.
Globally it is supported on the flavor branes and should therefore
not contribute to the Chern-Simons term \eqref{chernsim} on the 
GUT brane.
Therefore using the same conventions as in \cite{Blumenhagen:2008aw},
in this case the ansatz for the $SU(5)$ gauge field
strength reads
\bea
\label{fluxexpans}
    F&=&\sum_{a=1}^8  F^a_{SU(3)} 
        \left(\begin{matrix}  
             \lambda_a/2 & 0 \\
             0 & 0  
          \end{matrix}\right) 
        +\sum_{i=1}^3  F^i_{SU(2)} 
        \left(\begin{matrix}  
             0 & 0 \\
             0 & \sigma_i/2  
          \end{matrix}\right)+ \nonumber\\
     &&   {\textstyle \frac{1}{6}}\, F_Y 
          \left(\begin{matrix}  
             -2_{3\times 3} & 0 \\
             0 & 3_{2\times 2} 
          \end{matrix}\right)+
        {\textstyle \frac{1}{5}}\, \overline{f}_Y
          \left(\begin{matrix}  
             -2_{3\times 3} & 0 \\
             0 & 3_{2\times 2} 
          \end{matrix}\right)\, ,\nonumber
\eea
where $\lambda_a$ denote the eight traceless Gell-Mann matrices and
$\sigma_i$ the three traceless Pauli matrices.
The capital letters $F_G$ denote the four-dimensional gauge fields
and the small letters $\ov f$ the internal background fluxes.   
Now, inserting the expansion (\ref{fluxexpans}) into the Chern-Simons term
(\ref{chernsim}) and extracting the $F\wedge F$ terms, we eventually
find the three MSSM gauge couplings
\bea
\label{gaugekins}
  \frac{1}{\alpha_s} &=& M_*^4 \, {\rm Vol}(S) -\frac{4}{50}\, M_*^4 \ell_s^4 \int_{S}  c^2_1({\cal L}_Y), \nonumber \\ 
   \frac{1}{\alpha_w}&=& M_*^4\, {\rm Vol}(S) -\frac{9}{50}\, M_*^4 \ell_s^4 \int_{S}  c^2_1({\cal
      L}_Y), \\
     \frac{3}{5}\,  \frac{1}{\alpha_Y}&=& M_*^4\, {\rm Vol}(S) -\frac{7}{50}\, M_*^4 \ell_s^4 \int_{S} c^2_1({\cal L}_Y) \; 
\nonumber
\eea
in terms of the F-theory mass scale $M_*^4=g_s^{-1}\ell_s^{-4}$.
As expected there is no way that all corrections are the same but 
they do still satisfy the relation \eqref{relc}. Therefore, also
in F-theory GUTs the running of the gauge couplings can be reconciled
with this flux induced splitting via thresholds of the Higgs triplets.

Next let us discuss the right-handed neutrinos in our F-theory models. 
The role of the right-handed neutrinos is played by the modes in ${\bf 1_5}$ representation. In our example the corresponding index is $\chi_{\nu_r^c}= 5$. While a detailed analysis of the neutrino sector is beyond the scope of this work, we recall from \cite{Marsano:2009gv} that the appearance of a $U(1)_X$ is in conflict with the neutrino scenarios of \cite{Tatar:2009jk, Bouchard:2009bu}. Indeed the extra selection rule forbids Majorana masses for $\nu_R^c$ while at the same time allowing for unsuppressed order one Dirac masses in terms of the operator ${\bf \ov 5_m \, 5_H \, 1}$. 

The $U(1)_X$ selection rule for Majorana masses can be bypassed by non-perturbative effects as in the context of weakly coupled Type II models. In fact the generation of Majorana masses for $\nu_R^c$ by Euclidean D3-branes in Type IIB \cite{Blumenhagen:2006xt, Ibanez:2006da} has an analogue also in F-theory. Achieving an intermediate scale for the Majorana masses proportional to ${\rm exp}(-{\rm Vol}_{inst.}/g_s)$ involves even less fine-tuning of the instanton volume as compared to the weakly coupled cousin models since the dilaton need not be small in F-theory. 

Another possibility to avoid the selection rule for Majorana masses is to break $U(1)_X$ directly, as pointed out already in \cite{Tatar:2009jk}. In fact, as described in \secref{SUSY_sec} forming a non-split extension out of ${ V}$ and ${L}$ has exactly this effect. After all the extension moduli are nothing other than the ${\bf 1_5}$, i.e. the right-handed neutrinos. Giving a  VEV to the extension moduli automatically renders the ${\bf 1_5}$ massive. 

Alternatively, the two neutrino scenarios of ref.~\cite{Bouchard:2009bu} are based on a $U(1)_{PQ}$ symmetry with charge assignment
\beq
  {\bf 10_1}, \qquad \bf (5_H)_{-2}, \qquad \bf (\ov 5_H)_{-3}, \qquad \bf (\ov 5_m)_{2} .
\eeq
We would like to point out that this charge assignment can in principle be achieved also in the context of the $S[U(4) \times U(1)]$ spectral cover considered in this paper. To this end one has to flip the role of
\beq
  {\bf \ov 5_m } \leftrightarrow {\bf \ov 5_H}. 
\eeq
Indeed the Yukawa couplings ${\bf 10 \, 10 \, 5_H}$ and   ${\bf 10 \, \ov 5_m \, \ov 5_H}$ are still allowed while the dimension-4 proton decay operators are forbidden. In the present form of the split spectral cover this means that the ${\bf \ov 5_m}$ and $\bf  5_H$ are localized on what we called ${\cal P}_H$ while the $\bf \ov 5_H$ emerges from the curve called ${\cal P}_m$. Another advantage of this scenario beyond the neutrino sector is that the missing partner mechanism for $H_u$ and $H_d$ forbids dimension-5 proton decay.  At the level of chiral indices an explicit realization of such a scenario requires that
\beq
  \int_{{\cal P}_{H}} \gamma_u + \frac14 \pi_4^* \zeta  = 2, \qquad \int_{{\cal P}_{m}}  \gamma_u + \frac14 \pi_4^* \zeta - \pi_1^* \zeta   = 1.
\eeq
Clearly the analysis of the vector-like spectrum on ${\cal P}_H$ is even more crucial now. Also it must be ensured that the ${\bf 5_H}$ and ${\bf \ov 5_m}$ which appear as vector-like pairs on $P_H$ do not acquire high-scale masses by pairing up. To remedy this once again  factorization of $P_H$ is required. Solving this challenge is therefore of importance both for the Higgs sector and neutrino physics at the same time. A more refined analysis is also required because in a split spectral cover with maximal monodromy group and with $\bf {5}_H$ and $\bf { \ov 5}_H$ localised on distinct curves, extra light exotics will appear \cite{Marsano:2009gv}.

\section{Conclusions}\seclabel{sec_concl}

In this paper we have constructed a global F-theory $SU(5)$ GUT model on a compact Calabi-Yau fourfold which has three chiral matter generations due to a consistent treatment of the gauge flux using the spectral cover approach.  We have kept the overall presentation of the necessary tools and techniques rather general so they are easily applicable to further model searches in this direction. While our models are genuine F-theory compactifications which do not admit a simple orientifold description, the methods to construct the geometries and gauge fluxes have been inspired by the Type IIB orientifold models \cite{Blumenhagen:2008zz} and our work on their F-theory uplifts \cite{Blumenhagen:2009up}.

Our approach to construct the Calabi-Yau fourfold geometry can be summarized as follows. The starting point is a very ample Fano threefold base, which guarantees the existence of a non-singular elliptically fibered Calabi-Yau fourfold. If the base is a hypersurface the Calabi-Yau fourfold is a complete intersection of the base constraint and the Weierstrass model. The del Pezzo surface supporting the GUT brane is then obtained via a del Pezzo transition in the base. We have found that the existence of a {\bf 10} matter curve requires the GUT seven-brane to wrap a non-generic del Pezzo surface which can only be shrunk to a curve instead of a point, and hence arises from a del Pezzo transition blowing up a singular curve in the base. 
We have constructed the singular Calabi-Yau fourfold with the appropriate $SU(5)$ degeneration of the elliptic fiber as a complete intersection of hypersurfaces using toric geometry. Toric geometry also allows us to explicitly resolve the $SU(5)$ singularities and study the geometry of the resulting Calabi-Yau fourfold. 

Inspired by earlier work on models with heterotic dual we have utilized the spectral cover approach to describe the gauge flux necessary for the generation of chiral matter. 
By embedding an additional $U(1)_Y$-hypercharge flux into $SU(5)$ we have then broken the GUT group to the Standard Model. In order to avoid vector-like exotics we have found that a further twist of the $U(1)_Y$-bundle similar to the construction detailed in \cite{Blumenhagen:2008zz} is required.
Furthermore, the hypercharge flux is restricted by the requirement that the associated $U(1)_Y$ gauge boson does not acquire a mass via the St\"uckelberg mechanism \cite{Donagi:2008kj}. 

As discovered in \cite{Tatar:2009jk} and further worked out in \cite{Marsano:2009gv} a splitting of the spectral cover is necessary to ensure a proper localization of $\mathbf{\ov 5}_m$ and $\mathbf{5}_H+\mathbf{\ov 5}_H$ states on distinct matter curves. This prohibits proton decay by avoiding the generation of the $\mathbf{10}\,\mathbf{\ov 5}_m\,\mathbf{\ov 5}_m$ Yukawa coupling and subsequent dangerous dimension-4 operators. Our main focus has been on the correct treatment of gauge flux in such a scenario. As an important finding we have  identified the gauge bundle on the matter branes for the split spectral cover as an $S[U(4)\times U(1)_X]$ bundle. This is the direct analogue of the construction developed previously in the heterotic context in \cite{Blumenhagen:2005ga, Blumenhagen:2006ux, Blumenhagen:2006wj, Weigand:2006yj}. By applying the gauge flux quantization conditions we find that 
one is generically forced to turn on the universal gauge flux, i.e.~a non-trivial gauge flux is actually required for overall consistency. Using this overall construction we have been able to bypass the 'three chiral generation' no-go theorem of \cite{Donagi:2009ra} for 'universal gauge flux'. The appearance of a massive $U(1)_X$ induces a field-dependent Fayet-Iliopoulos term and ensuring D-flatness fixes one linear combination of K\"ahler moduli and $U(1)_X$ charged matter --- a constraint that has to be taken into account in the construction of consistent vacua.

Having established the geometry and flux setting we have analyzed the global consistency conditions explicitly, which are out of reach for any local model. While the seven-brane tadpole is automatically satisfied in F-theory models if the compact fourfold is of Calabi-Yau type, the three-brane tadpole is known to receive  contributions both from the curvature and the gauge flux. Inspired by heterotic/F-theory duality we have proposed a remarkably simple method to compute the Euler characteristic for a certain class of singular four-folds. We have checked by comparison both with the result of a brute-force toric computation and of the method of \cite{Andreas:2009uf, Curio:1998bva} that this yields the correct value even for models without a heterotic dual. This lends further credibility to the applicability of the spectral cover construction in such situations.

We have then applied the aforementioned steps to an explicit model using the Fano threefold $\bbP^4[4]$ and a geometric transition along a singular $\bbP^1$ to obtain a non-generic $dP_7$ divisor for the GUT brane. We have shown that it is indeed possible to find a compact global $SU(5)$ GUT model with three chiral matter generations and the phenomenologically relevant ${\bf 10 \, 10 \, 5_H}$ and ${\bf 10 \, \ov 5_m \, \ov 5_H}$ Yukawa couplings while the split spectral cover prevents dangerous dimension-four proton decay.

Clearly, more work is needed to improve on the phenomenology of this model. In this article, we have not evaluated the explicit cohomology groups required to determine the exact matter spectrum, but only computed the chiral index. As discussed in \cite{Marsano:2009gv} another urgent task is to describe the localization of $H_u$ and $H_d$ on separate curves within the spectral cover approach. This is required to avoid unacceptably high $\mu$ terms and, in fact, also to suppress dimension-five proton decay. In any case our treatment of the gauge flux is expected to be relevant also on such more refined spectral covers. As far as the global properties of our particular model are concerned we have observed an overshooting of the three-brane tadpole condition.
However, in view of the extremely simple nature of the used geometry we are confident that a more extensive model search following the presented outline should provide an abundance of globally consistent models without overshooting. This expectation is corroborated by the general tendency of an increasing fourfold Euler characteristic for Weierstrass models on threefolds with larger Hodge numbers. 

Eventually one hopes to combine the successful F-enomenology of the approach of \cite{Donagi:2008ca, Beasley:2008dc, Beasley:2008kw, Donagi:2008kj} with full stabilization of the geometric moduli, possibly similar in spirit to \cite{Braun:2008ua, Braun:2008pz}. While F-theory/Type IIB constructions indeed seem the right corner of the string landscape also for this latter undertaking, all explicit constructions with stabilized moduli so far contain no or only trivially unrealistic particle physics. As pointed out in \cite{Blumenhagen:2007sm} and further elaborated on in \cite{Blumenhagen:2009gk}, these two questions can actually not be considered independently, and the interplay of the closed and open sector is crucial to understand such pressing questions as the scale of supersymmetry breaking and its mediation in consistent models.

\subsection*{Acknowledgements}
We gratefully acknowledge discussions with C. Cordova, J.~Heckman, H.~Jockers, S.~Kachru, A.~Klemm, W.~Lerche, M.~Wijnholt, and E.~Witten.
TW is particularly grateful to S.~Kachru for constant encouragement.
TG and TW would like to thank the KITP Santa Barbara, and the MPI Munich for hospitality during the preparation of this work. 
TG also gratefully acknowledges hospitality and support from the Simons Center for Geometry and Physics, Stony Brook.
TW is supported by the DOE under contract DE-AC03-76SF00515.
This work was supported in parts by the 
European Union 6th framework program MRTN-CT-2004-503069 ``Quest for unification'', 
MRTN-CT-2004-005104 ``ForcesUniverse'', MRTN-CT-2006-035863 ``UniverseNet'',
SFB-Transregio 33 ``The Dark Universe'' by the DFG.

\appendix

\section{Computation of the 3-brane tadpole}\applabel{Tadpoles_sec}

As pointed out in \secref{GeomTad} both the curvature terms and the gauge flux on the 7-branes lead to induced 3-brane charge. In a compact setting its cancellation generically requires the introduction of 3-branes, whose number is constrained by~\cite{Sethi:1996es}
\beq \label{D3appendix}
  N_{3} = \frac{\chi(Y)}{24} - \frac12 \int_Y G\wedge G . 
\eeq
Recall from \secref{GeomTad} that special care has to be taken in computing the Euler characteristic of the singular Calabi-Yau fourfold $Y$. The elliptic fourfold $Y$ will become singular if the gauge enhancement over the divisor $S$ in the base $B$ is encoding a non-Abelian group $H$. In this appendix we present the details that lead us to formula \eqref{chi-prop} for $\chi(Y)$. 

Let us first consider F-theory models that admit a heterotic dual. More precisely, we assume that the elliptic fourfold $Y$ admits also K3-fibration with base $B_2$. This means that the base $B$ of the elliptic fibration has itself the structure of a $\mathbb P^1$ fibration with base $B_2$. We demand that $B_2$ is the divisor $S$ over which the elliptic fiber degenerates to an enhanced gauge symmetry. This corresponds to the special case in which the ALE fibration over $S$ is defined globally, not just locally as in general models. F-theory on such $Y$ is dual to the heterotic $E_8 \times E_8$ string on a Calabi-Yau threefold $Z$ elliptically fibered over $B_2$. A gauge group $H$ on $S$ is engineered by embedding two bundles $V_1$ and $V_2$  of structure group  $G$ and $E_8$ into the respective factors of $E_8 \times E_8$ such that $H = E_8^{(1)}/ G$. For $G=SU(N)$ the bundle $V_1$ is described by a spectral cover on $Z$ in a similar way as reviewed in \secref{gauge_fluxSU(5)}. For bundles of exceptional structure the actual procedure is somewhat different \cite{Friedman:1997yq}, but in either case the bundles $V_i$ are both determined by an element $\eta_i \in H^2(Z; \mathbb Z)$ with
\beq \label{eta12}
  \eta_1 = 6 c_1(S) - t\ , \qquad \eta_2 = 6 c_1(S) + t\ .
\eeq

In order to make contact to \eqref{D3appendix} we use the key observation \cite{Friedman:1997yq} and identify the number $N_{3}$ of three-branes in \eqref{D3appendix} with the number of M5-branes in the heterotic string. Anomaly cancellation in the heterotic string requires the inclusion of M5-branes, whose number is given, for the simple embedding considered above, by
\beq \label{M5}
  N_{\mathrm{M5}} = \int_{B_2}\, c_2(Z) - c_2(V_1) - c_2(V_2).
\eeq
The second Chern classes of $V_1$ and $V_2$ have been computed in \cite{Friedman:1997yq} as
\beq \label{c2V}
  \bal
    & \int_{B_2} c_2(V_1) = \int_{B_2} \eta_1 \sigma - \frac{1}{24}  \chi_{SU(n)^{(1)}} - \frac{1}{2} \int_{B_2} \pi_{n*}(\gamma^2), \\
    & \int_{B_2} c_2(V_2) = \int_{B_2} \eta_2 \sigma - \frac{1}{24} \chi_{E_8^{(2)}},
  \eal
\eeq
with the expressions for $\chi_G$ displayed in \tableref{chigaugegroups} in \secref{GeomTad}. The superscripts ${}^{(1)}, {}^{(2)}$ remind us to use the values of $\eta_i$ for the two respective bundles
as in \eqref{eta12}. The second Chern class of $Z$ has also been determined in \cite{Friedman:1997yq} as
\beq
  c_2(Z) = 12 \sigma c_1(B_2) + 11 c_1^2(B_2) + c_2(B_2).
\eeq

By construction the $\sigma$ dependent pieces in \eqref{M5} cancel. Identifying $N_{\mathrm{M5}}=N_3$ \cite{Friedman:1997yq} one finds that
\beq \label{D3b}
  N_{3} = \int_{B_2} \big( 11 c_1^2(B_2) + c_2(B_2) \big) +   \frac{1}{24} \left( \chi_{SU^{(1)}(n)}   +   \chi_{E_8^{(2)}} \right) +  \frac{1}{2} \int_{B_2} \pi_{n*}(\gamma^2).
\eeq
Given the relation between the gauge flux and the form $\gamma$ in F-theory, it is furthermore natural to conclude $\int G \wedge G = - \int \pi_{n*}(\gamma^2)$ as in \eqref{Gflux} \cite{Curio:1998bva}.
The terms in brackets must then equal $\chi(Y)$ in \eqref{D3b}. For heterotic/F-theory dual pairs $Y, Z$ this follows directly by duality. 

Our claim is that this is actually true more generally and thus serves as a simple and direct means to compute the Euler characteristic of a singular $Y$. To this end one must express the first two terms in \eqref{D3b} directly in geometric invariants of $Y$ without any reference to a K3-fibration. For F-heterotic dual pairs we know that if we replace $\smash{\chi_{SU(n)^{(1)}}}$ by $\smash{\chi_{E_8^{(1)}}}$, the gauge group is completely broken on the heterotic side, which corresponds to a \emph{non-singular} fourfold K3-fibered over $B_2$ with corresponding 
\beq \label{chia_app}
 \chi^*(Y) =  24 \int_{B_2} \big( 11 c_1^2(B_2) + c_2(B_2) \big) +  \chi_{E_8^{(1)}}  +   \chi_{E_8^{(2)}}\ .
\eeq
Eliminating $\chi_{E_8^{(2)}}$ leads to
\beq \label{chi-prop_app}
  \chi(Y) = \chi^*(Y) + \chi_{SU^{(1)}(n)} - \chi_{E_8^{(1)}}. 
\eeq
Our conjecture is that this holds for every Calabi-Yau fourfold $Y$ which is elliptically fibered over a base with the following property: It must also allow for a Weierstrass model whose discriminant consists only of $\mathrm{I}_1$ or $\mathrm{II}$ components but contains no loci of further non-Abelian enhancements. Here $\chi^*(Y)$ denotes the Euler characteristic of this $\mathrm{I}_1$ Weierstrass model. 
For  $\chi^*(Y)$  a general expression was derived in \cite{Sethi:1996es},
\beq \label{klemmyau_app}
  \chi^*(Y)=  12 \int_B c_1(B)\, c_2(B) + 360 \int_B c^3_1(B)\ .
\eeq




\clearpage
\bibliography{rev40}  
\bibliographystyle{utphys}

             
\end{document}